%% file: vou_blazar.tex
\documentclass[final,3p,times,twocolumn,authoryear]{elsarticle}
\journal{Astronomy and Computing}

\usepackage{geometry}
\usepackage{amsmath} 
\usepackage{graphicx,rotating}
\usepackage{threeparttable}
\usepackage{longtable} 
\usepackage{graphicx}
\usepackage{amsmath}
\usepackage{amssymb}
\usepackage[compact]{titlesec}
\usepackage{rotating,booktabs}
\usepackage{hyperref}
\usepackage{movie15}
\usepackage{booktabs,longtable,threeparttable}
\usepackage{multicol}
\usepackage{indentfirst}
\usepackage{epstopdf}
\usepackage{subfig}
\usepackage{enumerate}
\usepackage[inline]{enumitem} 
\usepackage{caption}
\usepackage{color}
\usepackage{natbib}
\usepackage{url}
\usepackage{array}
\usepackage[colorinlistoftodos,prependcaption]{todonotes}
\usepackage{lscape}
\usepackage[normalem]{ulem}

\input{commands}

\begin{document}

\begin{frontmatter}


\title{The Open Universe VOU-Blazars tool}

\fntext[label1]{Tsung-Dao Lee Institute, Shanghai Jiao Tong University, 800 Dongchuan RD. Minhang District, Shanghai, China}
\fntext[label2]{ICRANet, Piazza della Reppublica 10, Pescara, Italy}
\fntext[label3]{Jacobs University, Physics and Earth Sciences, Campus Ring 1, 28759, Bremen, Germany}
\fntext[label4]{Agenzia Spaziale Italiana, Via del Politecnico s.n.c., Rome, Italy}
\fntext[label5]{Institute for Advanced Study, TUM, Lichtenberg strasse, Garching, Germany}

%\tnotetext[label1]{alpha-version}
 \author{Yu-Ling Chang\textsuperscript{*} \fnref{label1,label2}}
 \ead{ylchang@sjtu.edu.cn}
%  \cortext[cor1]{Corresponding author:}
 %\address{Address\fnref{}}
 %\fntext[]{}

\author[label2,label3]{Carlos H. Brandt}
\author[label4,label5,label2]{Paolo Giommi}

\input{abstract}

\end{frontmatter}

%\listoftodos


\input{introduction}

\input{run}

\input{table}

\input{results}

\input{application}

\input{conclusion}

\input{acknowledgment}

\input{reference}

%\todo[inline,color=green]{Refine the citation...}




%\begin{thebibliography}{00}



%\end{thebibliography}
\end{document}

%% file: commands.tex
\newcommand{\gr}{{$\gamma$-ray}}
\newcommand{\OU}{Open Universe}
\newcommand{\lsim}{{\lower.5ex\hbox{$\; \buildrel < \over \sim \;$}}}
\newcommand{\gsim}{{\lower.5ex\hbox{$\; \buildrel > \over \sim \;$}}}
\newcommand{\nupeak}{$\nu_{\rm peak}$}

\newcommand{\foothref}[1]{\footnote{\url{#1}}} % for URLs in footnotes
\hypersetup{draft}

%% file: abstract.tex
\begin{abstract}
\textit{Context:} Blazars are a remarkable type of Active Galactic Nuclei (AGN) that are playing an important and rapidly growing role in today's multi-frequency and multi-messenger astrophysics.
%The property that makes them unique, among other AGN, is the presence of a relativistic jet that happens to be pointing in the direction of the Earth. This is a very special geometrical situation that causes the non-thermal radiation emitted by the energetic particles inside the jet to be strongly amplified and appear to be highly and rapidly variable.
In the past several years, blazars have been discovered in relatively large numbers in radio, microwave, X-ray and $\gamma$-ray surveys, and more recently have been associated to high-energy astrophysical neutrinos and possibly to ultra-high energy cosmic rays.
Blazars are expected to dominate the high-energy extragalactic sky that will soon be surveyed by the new generation of very high-energy $\gamma$-ray observatories such as CTA.
%as well as  playing a central role in the emerging field of multi-messenger astrophysics.
%as these sources 
In parallel to the discovery of many blazars at all frequencies, the technological evolution, together with the increasing adoption of open data policies, is causing an exponential growth of astronomical data that is freely available through the network of Virtual Observatories (VO) and the web in general, providing an unprecedented potential for multi-wavelength and multi-messenger data analysis.

\textit{Product:} We present \textit{VOU-Blazars}, a tool developed within the \OU\, initiative that has been designed to facilitate the discovery of blazars and build their spectral energy distributions (SED) using public multi-wavelength photometric and spectral data that are accessible through VO services.
%\textit{VOU-Blazars} was  an initiative under the auspices of the United Nations with the objective of largely improving the availability and the use of space science and astronomy data. 
The tool is available as source code, as a Docker container, and as a web-based service accessible within the \OU\ portal\foothref{http://openuniverse.asi.it}.

\textit{Methods:} VOU-Blazars implements a heuristic approach based on the well known SED that differentiate blazars from other astronomical sources.
The VOU-Blazars outputs are flux tables, bibliographic references, sky plots and SEDs.
%suitable for further analysis by the users.

\textit{Results:} This paper describes the working mechanism of the tool, gives details of the catalogues and on-line services that are used to produce the output, and gives some examples of usage.
 VOU-Blazars has been extensively tested during the selection of new high-energy peaked (HSP) blazars recently published in the 3HSP catalog, and has been used to search for blazar counterparts of Fermi 3FHL, Fermi 4FGL, AGILE \gr\ sources, and of IceCube astrophysical neutrinos. 

% To further examinate the counterpart for them, we built a tool called VOU-Blazars to find blazar candidates using virtual observatories(VO). 
% There were 60 catalogs, from radio to \gr\, applied in VOU-Blazars. 
% The tool could figure out all the possible blazars candidates from all available radio or X-ray sources in a extensive area. 
% By assigning a specified coordinate, the VOU-Blazars will return all potential blazars and will further show the spectral energy distribution (SED) of the interested source. 
% Till now, the tool have been applied to find Fermi 3FHL and FL8Y counterparts, and also the IceCube neutrinos events counterpart. 
% Apart from that, some of the new high synchrotron peaked blazars reported in the 3HSP catalog are found with the VOU-Blazars. 
\end{abstract}

%% file: introduction.tex
\section{Introduction}
\label{brief}
We present a software tool developed in the context of the Open Universe Initiative \citep{ougiommi} that has been designed to facilitate the identification and the broad-band spectral study of blazars and other astrophysical sources, based on public multi-frequency data retrieved from the Virtual Observatories\footnote{http://www.ivoa.net/} network. 

Blazars are a special type of Active Galactic Nuclei (AGN) distinguished by the emission of strong non-thermal radiation across the entire electromagnetic spectrum \citep{Padovani2017}, a unique characteristics among extragalactic objects that gives them a central role in contemporary $\gamma$-ray and multi-frequency astronomy, and likely in the emerging field of multi-messenger astrophysics.  
Developing efficient tools useful for the detailed study of these sources is therefore important and timely to support a rapidly evolving research in this field.

\begin{figure}[h!]
\begin{center}
\includegraphics[width=1.\linewidth]{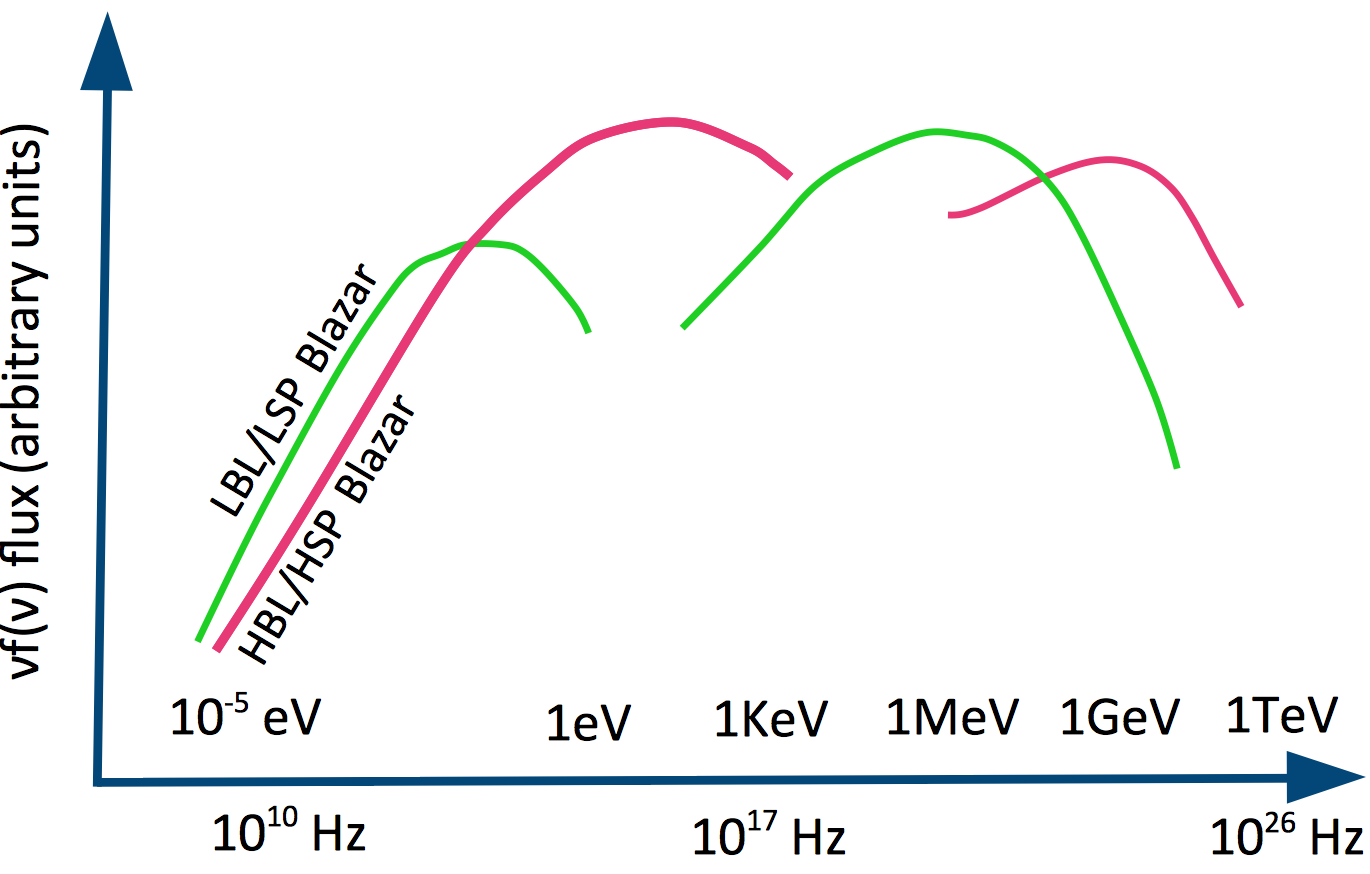}
\end{center}
\caption{The SED of different types of blazars (adapted from \citet{Padovani2017}). See text for details.}
\label{SEDs}
\end{figure}

From a physical viewpoint the non-thermal continuum seen in the spectral energy distribution (SED) of blazars is due to the radiation emitted by energetic particles moving in a relativistic jet emerging from the central super-massive black hole\citep{Padovani2017}. %accreting matter in the center of a galaxy.

From the observational viewpoint, the multi-waveband SED plot (Fig.~\ref{SEDs}) of any blazar well covered from radio to high-energy (to x-ray, $\gamma$-ray) typically presents two distinct features: high variability and two large bumps.
The first bump is generally attributed to synchrotron emission, while the second one may be due to inverse Compton scattering or to radiation originating in a hadronic or lepto-hadronic scenarios \citep{Reimer2012,Cerruti2011,Petropoulou2015,Petropoulou2016,Gokus2018}.
%Variability is a consequence of fluctuations in the underlying accretion process, we will mostly likely see vertical variations -- \textit{i.e.}, flux -- in our SED; Variability periods may vary from days to years and affect the whole spectrum, it is particularly flagrant around the bumps' peaks -- indicating that the physical mechanisms responsible for the lower energy bump ($10^{12} < Hz < 10^{16}$ and the higher energy bump ($Hz > 10^{18}$) are effectively related. 

%The position where the first bump peaks has been used to further classify the blazars family in low synchroton peaked (LSP or LBL), and high synchroton peaked (HSP or HBL) objects.\citep{Padovani1995,Abdo2010}, reflecting the maximum energy of the radiating particles (see Fig.\ref{SEDs}). Objects peaking at intermediate energies are called ISP (or IBL) blazars.
The position where the \textit{first} bump peaks has been used to further classify blazars into low synchrotron peaked (LSP) and high synchrotron peaked (HSP) objects\citep{Padovani1995,Abdo2010}. LSP blazars will present an associated peak energy below $Hz < 10^{14}$, while HSP blazars' peak at $Hz > 10^{15}$.
Objects peaking at intermediate energies are called ISP blazars.

%\sout{The radiation from the nuclear region is mostly non-thermal and originates from the accretion onto a supermassive black hole and, in large part, within a jet of relativistic matter that happens to be pointing close to the direction of the Earth. Because of their nature and of the particular geometrical situation these objects are by far the most common type of extragalactic sources so far detected in the microwave and in the \gr\, sky.}
%\sout{Building ools to locate and identify Blazars based on the multi-frequency data available is necessary to match the needs of present and future astronomical research.} 
%to select blazars candidates from a list of objects in areas of the sky specified by the user.
%\sout{The tool surveys multi-wavelength catalogs using Virtual Observatory services, dynamically correlates the sources in different energy bands to locate blazars candidates, and builds individual spectral energy distributions
%to in-code knowledge is applied 
%to verify the blazar nature and assign a blazar sub-class to each candidate.}
%\sout{The final word though is given by the user: }

Observations have shown that HSP blazars are bright and variable sources of high-energy \gr s (TeVCat)\footnote{http://tevcat.uchicago.edu} and that they are likely the dominant component of the extragalactic very high-energy (VHE, E $>$ 100 GeV) background \citep{Padovani1993,Giommi2006,DiMauro2014,Giommi2015,Ajello2015}. 
Very high-energy emitters are rare and still poorly understood astrophysical sources. So far only a few hundred of them have been detected in the VHE band. %(See TeVCat). 
The identification of VHE sources is receiving increasing attention, especially in relation with the likely connection with high-energy neutrinos and ultra high-energy cosmic-rays. 
A non negligible fraction of the cataloged \gr\ or VHE sources, %such as {\it Fermi}, Cherenkov Telescope Array (CTA) source
or neutrino events, are not associated to lower energy sources yet. 
%Blazars as one of the few types of objects that are expected to emit VHE radiation, thus finding the counterpart for VHE detections may be more effective if we begin with checking blazars. 
%According to \cite{Padovani1993,Giommi2006,DiMauro2014,Giommi2015,Ajello2015}, HSP blazars are the dominant population in the extragalactic VHE sky. , and the most promising candidates for extragalactic neutrinos sources. 

%why build this tool
%How and where the extremely high-energy emissions come from is still an open question.
%There are more and more evidences suggesting that there might be a connection between blazars and high-energy emission/particle, especially neutrino events. 
The recent association between the astrophysical high-energy neutrino IceCube-170922 with the bright blazar TXS0506+056 \citep{Icecube2018a,Icecube2018b,Padovani2018} represents a strong mark, and possibly a smoking gun, in the search for the sources of ultra high-energy (UHE) particles, opening up the new era of blazar studies. 
%Motivating by the fact that most of the extragalactic objects detected so far above a few GeV are HSPs  (\cite{Giommi2009,Padovani2015,Arsioli2015,Fermi3FHL2017}, see also TeVCat)
Motivated by such evidences and in view of the increasing sensitivity of the next generation of VHE and multi-messenger observatories, we developed a tool, VOU-Blazars, to mine the existing and constantly increasing multi-wavelength databases to support the search of new blazars and study their broad-band spectral properties and their temporal behavior.
%The tool is called VOU-Blazars, and the code .
%to find more possible counterparts for the VHE soruces, a tool to quickly identify all the possible blazars or blazar candidates was built, called VOU-Blazars. 

%brief introduction
The name, VOU, is the combination of VO (virtual observatory) and OU (open universe) as VOU-Blazars has been developed in the framework of \OU\ \footnote{http://openuniverse.asi.it/} and is based on VO protocols\footnote{http://www.ivoa.net/}.  
\OU\, is an initiative conceived to make astronomical data more openly available and usable by the widest possible community. Initially proposed by Italy to the United Nations Committee on the Peaceful Uses of Outer Space (COPUOS) in 2016\foothref{http://www.unoosa.org/res/oosadoc/data/documents/2016/aac_1052016crp/aac_1052016crp_6_0_html/AC105_2016_CRP06E.pdf}
%\textcolor{blue}{[cite A/AC./105/2016/CRP.6} 
\OU\, is now being actively developed by a number of countries in coordination with the UN Office for Outer Space Affairs (UNOOSA).

%an initiative under the auspices of the United Nations Committee on the Peaceful Uses of Outer Space (COPUOS) and coordinated by the United Nations Office for Outer Space Affairs (UNOOSA), with the objective of largely increasing the availability and the usability of space science data,
%extending the potential of scientific discovery to new participants in all parts of the world. 

%The main purpose to build this tool is to search blazar candidates effectively at certain position within a specified area based on searching with various multi-frequency VO catalogs to point out all the possible counterpart for VHE observations. 
The VOU-Blazar software is available as source code\foothref{https://github.com/ecylchang/VOU_Blazars}, as a tool encapsulated in a Docker container \foothref{https://hub.docker.com/r/chbrandt/voublazars/},
\foothref{https://www.docker.com}, and as a web service that can be activated from the \OU\ portal.
%The source code can be found at .
The databases used in the tool are all publicly available within the VO network.

%%%%%%%%%%%%%%%%%%%%%%%%%%%%%%%
%The tool is still under developing and could be download from \url{https://github.com/ecylchang/VOU_Blazars}.

%% file: run.tex
\section{Workflow}
\label{how} 
\begin{figure*} [htb!]
\begin{center}
\includegraphics[width=1.0\linewidth]{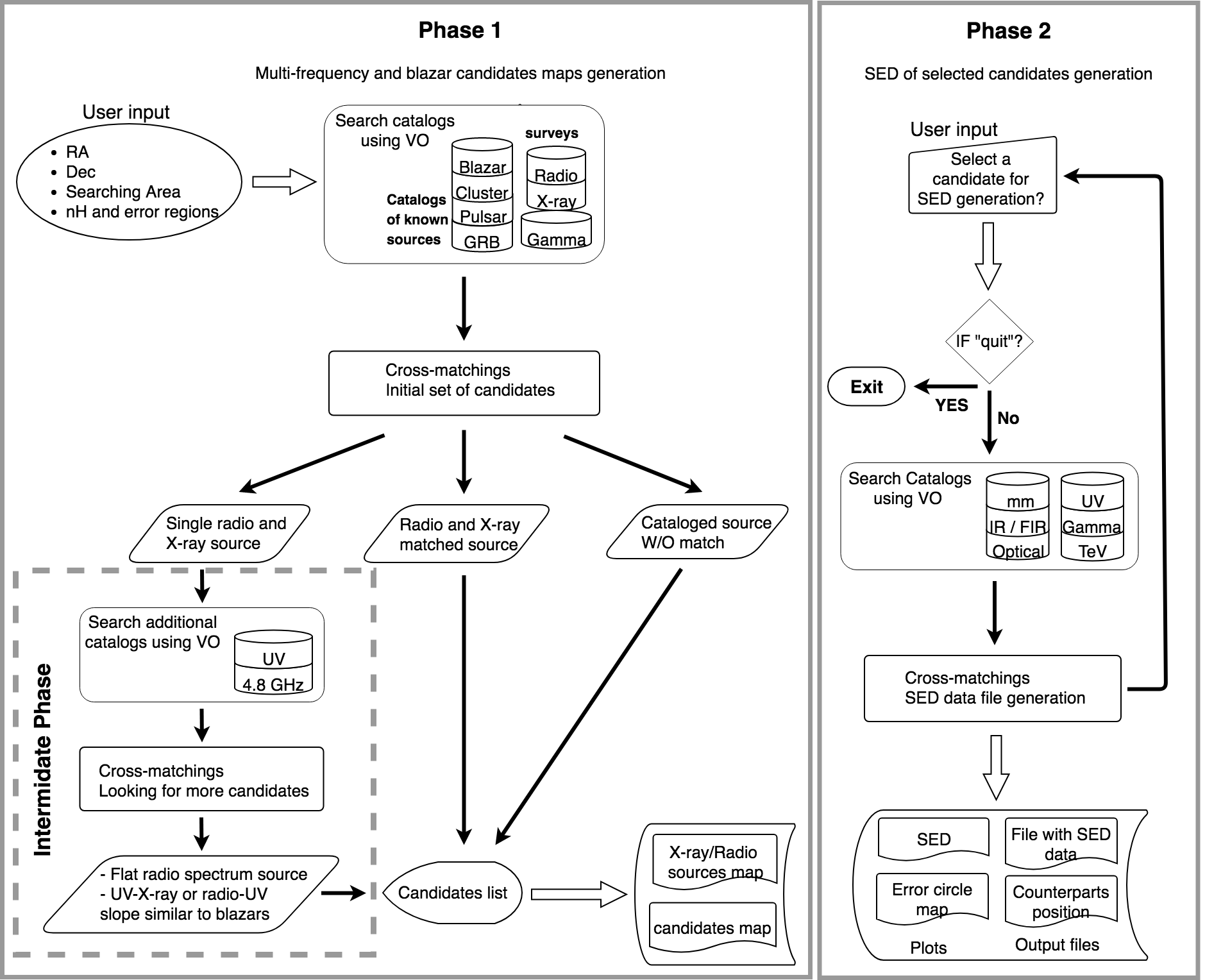}
\end{center}
\caption[Simple scheme of the working mechanism of the VOU-Blazars tool]{Simple scheme of the working mechanism of the VOU-Blazars tool.}
\label{scheme}
\end{figure*}
%%Applying MW data, introduce 2 phase.

It is well known since the early days of multi-frequency astronomy that blazars can be discovered through the analysis of radio, optical/IR and X-ray data (see e.g. \citet{Padovani1995,Perlman1998,Giommi1999,Chang2017,Arsioli2015}).
The VOU-Blazars software exploits this property as shown in scheme of Fig.~\ref{scheme}, which
illustrates the two-phase working mechanism of the tool. 
The first phase locates blazars candidates in the requested area, which can be relatively large, while the second phase deals with only one candidate at a time, generating and plotting the SED and the corresponding error circle map.
The error circle map is a plot where all the nearby detections, as well as their position error circle, are visualized.

%why radio and X-ray data first
The main purpose of the first phase is to identify blazar candidates within a given searching area, whose size that can be as small as one arcminute or less to several degrees.  
Given that the majority of the blazars are detected in both the radio and X-ray band, VOU-Blazars begins with searching for sources in all available lists of radio or X-ray emitters within the specified searching region.  
In this phase the tool retrieves data from several catalogs providing photometric data through VO conesearch pipeline \footnote{https://github.com/chbrandt/eada/tree/master/eada}, and estimates the possible presence of blazar candidates. 
%the most effective way to find blazar candidates is beginning with multi-frequency data.

%%%other catalogs
The tool also interrogates type-specific catalogs of known sources (blazars, \gr\ burst (GRB), pulsars, \gr\ sources, and clusters of galaxies) to further help the user on possible ambiguities. 
In the first phase, currently a total of 34-catalog searches are made through the conesearch pipeline. 
%We will call the flux values of \gr\ catalogs and CRATES catalog later in the second phase.
The catalogs used in this phase are listed in Table~\ref{vocat1}, while Table~\ref{band1} gives the frequencies of the X-ray data from each catalog.
%Here we applied these catalogs to see if there are \gr\ sources or flat radio sources nearby the candidates, and we obtained the flux value of them. 
%Apart from that, there were other blazars, pulsar, /gr/, and cluster catalogs are called. 

\begin{table*} [h!]
\begin{center}
\begin{threeparttable}
\begin{tabular}{l|l}
\hline\hline
Radio/Fermi Catalogs&X-ray Catalogs\\
\hline
NVSS \citep{Condon1998}&XMMSL Dr2 Clean \citep{Saxton2008}\\
FIRST \citep{White1997,Helfand2015}&3XMM Dr8 \citep{Rosen2015}\\
SUMSS V2.1 \citep{Manch2003}&RASS 2RXS \citep{Boller2016}\\
CRATES \citep{Healey2007}&WGACAT2 \citep{White2000}\\
Fermi 4FGL \citep{Fermi4FGL} &Swift 1SXPS \citep{Evans2014}\\
Fermi 3FHL \citep{Fermi3FHL2017}&SDS82\citep{sds}\\
Fermi 3FGL  \citep{Acero2015}&Einstein IPC \citep{Harris1993}\\
1BIGB SED \citep{Arsioli2018}&Einstein IPC slew survey \citep{Elvis1992}\\
MST-9Y \citep{Campana2018} &BMW-HRI \citep{Panzera2003}\\
FermiMeV \citep{Principe2018} &Chandra\citep{Evans2010}\\
2AGILE  \citep{Bulgarelli2019} &Swift OUSXB \citep{GiommiOUSXB}\\
 & Swift OUSXG \citep{Giommi2019a} \\
\hline\hline
Blazar/Pulsar/GRB/Quasar Catalogs&Cluster of galaxies Catalogs\\
\hline
5BZCat \citep{Massaro2015}&ZW CLUSTERS \citep{Zwicky1968}\\
3HSP\citep{Chang2019}&PLANCK SZ2 \citep{PlanckCollaboration2016b}\\
ATNF PULSAR \citep{Manchester2005}&ABELL \citep{Abell1989}\\
{\it Fermi} 2PSR \citep{TheFermiLATCollaboration2013}&MCXC \citep{Piffaretti2011}\\
Fermi GRB \citep{Ajello2019}&SDSS WHL \citep{Wen2009}\\
Million Quasar \citep{Flesch2015}&SW XCS \citep{Liu2015}\\
\hline\hline
\end{tabular}
%\begin{tablenotes}
% \item[a] Swift DeepSky data over Stripe82, http://vo.bsdc.icranet.org/sds82/q/cone/info
%\end{tablenotes}
\caption{Catalogs applied in the first phase}
\label{vocat1}
\end{threeparttable}
\end{center}
\end{table*}

\begin{table*} [h!]
\begin{center}
\begin{tabular}{llc}
\hline\hline
X-ray catalogue&Energy band&Energy of nufnu flux \\
& & for SED plotting\\
& & (keV) \\
\hline
XMMSL Dr2 Clean&b8 (0.2  - 12 keV)& 1.0  \\
 &b6 (0.2 - 2 keV)& 1.1  \\
 &b7 (2 - 12 keV) & 7.0  \\
3XMM Dr8&EP8 (0.2 - 12 keV)& 1.0  \\
 &EP1 (0.2 - 0.5 keV)& 0.35  \\
 &EP2 (0.5 - 1 keV)& 0.75  \\
 &EP3 (1 - 2 keV)& 1.5  \\
 &EP4 (2 - 4.5 keV)& 3.25  \\
 &EP5 (4.5 - 12 keV)& 8.25  \\
RASS 2RXS&0.1 - 2.4 keV& 1.0  \\
WGACAT2 &0.24 - 2 keV& 1.0  \\
Swift 1SXPS& Total 0.3 - 10 keV& 1  \\
&Soft 0.3 - 1 keV& 0.65  \\
&Medium 1 - 2 keV& 1.5  \\
&Hard 2 - 10 keV& 6  \\
XRT DeepSky / SDS82& Total 0.3 - 10 keV & 3.0  \\
&Soft 0.3 - 1 keV& 0.5  \\
&Medium 1 - 2 keV& 1.5  \\
&Hard 2 - 10 keV& 4.5  \\
Einstein IPC / IPC slew survey& 0.2 - 3.5 keV& 1.0  \\
BMW-HRI& 0.1 - 2.4 keV& 1.0  \\
Chandra&Full band 0.5 - 7 keV& 1.0  \\
&Ultra Soft 0.2 - 0.5 keV& 0.35  \\
&Soft 0.5 - 1.2 keV& 0.85  \\
&Medium 1.2 - 2 keV& 1.6  \\
&Hard 2 - 7 keV& 4.5  \\
Swift OUSXB / OUSXG &Full band 0.3 - 10.0 keV& 3  \\
&Soft 0.3 - 1.0 keV& 0.5  \\
&Medium 1.0 - 2.0 keV& 1.5  \\
&Hard 2.0 - 10.0 keV& 4.5  \\
&Interpolation 0.5 and 1.5 keV & 1.0 \\
\hline\hline
\end{tabular}
\caption{List of X-ray catalogues, energy bands and SED energies}
\label{band1}
\end{center}
\end{table*}

%%The algorithm of difine the type of sources. 
It is well known that the radio to X-ray flux ratios are very different in HSP and LSP blazars \citep[e.g.][]{Padovani1995, Padovani2003,Padovani2007a}. 
VOU-Blazars uses the spectral slope between radio and X-ray, ($\alpha_{\rm 1.4 GHz-1 keV}$  or $\alpha_{\rm rx}$\footnote{$\alpha_{\nu1- \nu2}=-\frac{\log(f_{\nu1}/f_{\nu2})}{\log(\nu_1/\nu_2)}$}), to classify every radio-X-ray matching sources in the first phase. 
Some of the cataloged blazars and pulsars (see table~\ref{vocat1} for catalog's references) might not have both radio and X-ray detections. 
We list them among the selected sources during the first phase as well. 
Table~\ref{class} shows the details of classification of the radio-X-ray matching candidates. 
All the blazar candidates as well as the cataloged sources (5BZCat, 3HSP, CRATES) are available in the second phase for a thorough examination. 

\begin{table*} [h!]
\begin{center}
\begin{tabular}{lc}
\hline\hline
$0.43<\alpha_{\rm rx}<0.78$, $\log$ \nupeak$>15.5$&HSP candidate\\
$0.43<\alpha_{\rm rx}<0.78$, $\log$ \nupeak$<15.5$&ISP candidate\\
$0.78<\alpha_{\rm rx}<0.95$&LSP candidate\\
$\alpha_{\rm rx}<0.43$&non-jetted AGN candidate\\
$0.95 > \alpha_{\rm rx}$&Unknown\\
\hline\hline
\end{tabular}
\caption{Classification of the candidates}
\label{class}
\end{center}
\end{table*} 

Several LSP blazars are not detected in the X-ray band, as the X-ray emission of this type of objects is weak and their flux may easily be below the sensitivity of currently available observations. For similar reasons, HSPs with relatively high \nupeak\ values might not be detected in radio surveys.
Here we call radio sources without X-ray detections "single radio sources", and vice versa. 
%It is worth checking those single radio or single X-ray sources in order not to lose interesting blazar candidates. 
If the requested searching area is small, a process called intermediate phase is ``triggered'' to search for additional blazar candidates among single radio or X-ray sources.
%To avoid producing too many candidates, the VOU-Blazars checks only the sources inside the error area during this phase. 
If no sources without radio-X-ray matching is present inside the error area, the intermediate phase will not be called. 
%That is, for HSPs with very faint flux, might not be detected by current  surveyradio, as Fig.\ref{norrmiss} depicted, especially for those extreme \nupeak\ blazars. 

The intermediate phase makes use of the GALEX, PMN, and GB6 catalogs (see table~\ref{vocat2} for catalog's references) to search for possible UV emission and to estimate the radio spectral index.
The searching radius for VO conesearch in this step is set to same as the input radius, same as in the first phase. 
The GALEX data then are converted to monochromatic fluxes and de-reddened using \citet{Fitzpatrick1999} extinction rule. 

In the second phase, the user can complete the data analysis of one or more of the candidates found in the first phase building a complete SED using data from additional 34 catalogs, covering the high-frequency radio, microwave, far IR, IR, optical, UV, hard X-ray, \gr, and TeV bands. The list of catalogs used in the second phase is given in Table~\ref{vocat1}. 

As blazars are extragalactic sources data from IR, optical, and UV catalogs need to be corrected for Galactic extinction. To take into account of this we follow the extinction rule in \citet{Fitzpatrick1999}.
Effective wavelengths and zero-magnitude fluxes applied for every bands are listed in Table~\ref{band2}.  

%% file: table.tex
\begin{table*} [h!]
\begin{center}
\begin{threeparttable}
\begin{tabular}{lc}
\hline\hline
4.8/8.6 GHz catalogs&Search Radius\\
                    & (arc-seconds)\\
\hline
GB6 \citep{Gregory1996}&30 \\
GB87 \citep{Gregory1992}&30 \\
PMN \citep{Griffith1994,Griffith1995,Wright1994,Griffith1993,Wright1996}&30 \\
ATPMN \citep{McConnell2012}&15 \\
AT20G \citep{Murphy2010}&15 \\
NORTH20CM \citep{White1992}&120 \\
CRATES \citep{Healey2007}&15 \\
\hline\hline
Microwave/mm catalogs&Search Radius\\
                    & (arc-minutess)\\
\hline
WISH 352Å MHz \citep{DeBreuck2002}&0.25 \\
PCNT \citep{Planck2018} & 3 \\
ALMA \citep{Bonato2019}& 0.25\\
%PCCS2 44 GHz \cite{PlanckCollaboration2016a}&3 \\
%PCCS2 70 GHz \cite{PlanckCollaboration2016a}&3 \\
%PCCS2 100 GHz \cite{PlanckCollaboration2016a}&3 \\
%PCCS2 143 GHz \cite{PlanckCollaboration2016a}&3 \\
%PCCS2 217 GHz \cite{PlanckCollaboration2016a}&3 \\
%PCCS2 353 GHz \cite{PlanckCollaboration2016a}&3 \\
\hline\hline
Far IR / IR catalogs&Search Radius\\
                    & (arc-seconds)\\
\hline
SPIRE250 \citep{Schulz2017}&15 \\
SPIRE350 \citep{Schulz2017}&15 \\
SPIRE500 \citep{Schulz2017}&15 \\
2MASS \citep{Skrutskie2006}&10 \\
AllWISE \citep{Cutri2013}&10 \\
\hline\hline
Optical catalogs&Search Radius\\
                    & (arc-seconds)\\
\hline
%USNO-B1 \cite{Monet2003}&10 \\
SDSS Dr14 \citep{Blanton2017,Abolfathi2017}&10 \\
HST GSC2.3.2 \citep{Lasker2008}&10 \\
Pan-STARRS Dr1 \citep{Chambers2016}&10 \\
Gaia Dr1 \citep{Arenou2017,Prusti2016}&10 \\
\hline\hline
UV / X-ray catalogs&Search Radius\\
                    & (arc-seconds)\\
\hline
UVOT SSC 1.1 \citep{Page2014,Yershov2014}&15 \\
GALEX \citep{Morrissey2007,Bianchi2018}&15 \\
XMMOMSUSS 3 \citep{Page2012}&15 \\
XRT spectral data \citep{Giommi2015b} &15 \\
OUSpectrum \citep{Giommi2019a}& 15\\
BAT 105 Months \citep{Baumgartner2013}&10 acrmin\\
\hline\hline
\gr\ catalogs&Search Radius\\
                    & (arc-minutes)\\
\hline
Fermi 3FGL \citep{Acero2015}&20\\
Fermi 2FHL \citep{Ackermann2015}&20\\
Fermi 3FHL \citep{Fermi3FHL2017}&20\\
Fermi 4FGL \citep{Fermi4FGL}&20\\
1BIGB SED \citep{Arsioli2018}& 10\\
2AGILE \citep{Bulgarelli2019}& 50 \\
Fermi MeV \citep{Principe2018}& 30\\
\hline\hline
TeV/IACTs catalogs&Search Radius\\
                    & (arc-minutes)\\
\hline
MAGIC &10\\
VERITAS&10\\
\hline\hline
\end{tabular}
%\begin{tablenotes}
% \item[a] http://vo.bsdc.icranet.org/xrtspec/xrtspec/cone/info
% \item[b] https://fermi.gsfc.nasa.gov/ssc/data/access/lat/fl8y/
%\end{tablenotes}
\caption{Catalogs applied in the second phase}
\label{vocat2}
\end{threeparttable}
\end{center}
\end{table*}

\begin{table*} [h!]
\begin{center}
\begin{threeparttable}
\begin{tabular}{ccccl}
\hline\hline
Catalog&Band&Zero-magnitude flux&Effective wavelength&Reference\\
& & Jy & \AA &\\
\hline
2MASS&J&1594 &12350  &\cite{Skrutskie2006} \\
&H&1024 &16620  &\\
&K&666.7 &21590  &\\
AllWISE&W1&309.540  &34000  &\cite{Cutri2013} \\
&W2&171.787  &46000  & \\
&W3&31.674  &120000  & \\
&W4&8.363  &220000  & \\
USNO&B&4260  &4400  &\cite{Bessell1979}\\
&R&3080  &6400  &\\
SDSS&u&3631  &3568  &\cite{Doi2010}\\
&g&3631  &4653  &\\
&r&3631  &6148  &\\
&i&3631  &7468  &\\
&z&3631  &8863  &\\
GSC&U&1810  &3600  &\cite{Bessell1979}\\
&B&4260  &4400  &\\
&V&5500  &3640  &\\
&R&3080  &6400  &\\
&I&2550  &7900  &\\
PanSTARRs&g&3631  &4810  &\cite{Tonry2012}\\
&r&3631  &6170  &\\
&i&3631  &7520  &\\
&z&3631  &8660  &\\
&y&3631  &9620  &\\
GAIA&G&2918   \tnote{a}&6730  &\cite{Jordi2010}\\
UVOT&u&3631  &3501 &\cite{Poole2008}\\
&b&3631  &4329 &\\
&v&3631  &5402 &\\
&w1&3631  &2634 &\\
&m2&3631  &2231 &\\
&w2&3631  &2030 &\\
GALEX&FUV&3631  &1538.6 &\cite{Morrissey2007}\\
&NUV&3631  &2315.7 &\\
XMMOM&u&3631  &3440 &\cite{Page2012}\\
&b&3631  &4500 &\\
&v&3631  &5430 &\\
&w1&3631  &2910 &\\
&m2&3631  &2310 &\\
&w2&3631  &2120 &\\
\hline\hline
\end{tabular}
\begin{tablenotes}
 \item[a] The G band magnitude is measured in Vega system, it depends on the
flux densities of A0V star, and the zero magnitude flux here is obtained from A0V star template from STSDAS calibrated database system at \url{http://www.stsci.edu/hst/observatory/crds/calspec.html}.
\end{tablenotes}
\caption{Magnitude reduction details for catalogs used in the second phase}
\label{band2}
\end{threeparttable}
\end{center}
\end{table*}

\begin{table*} [h!]
\begin{center}
\begin{tabular}{lc|lc}
\hline\hline
Color&Meaning&Symbol&Meaning\\ 
\hline
Orange&HBL candidate&Open circle&X-ray component\\
Cyan &IBL candidate& Filled circle&Radio component\\
Dark Blue&LBL candidate& & \\
Green&Non-jetted AGN candidate&&\\
Black&Unknown type radio + X-ray source&&\\
\hline\hline
Symbol and Color&Meaning&Symbol and Color&Meaning\\
\hline
Gold Star&3HSP source&Red filled Circle&radio source\\
Gold Diamond&5BZCat source&Blue open circle&X-ray source\\
Question Mark&Cluster of galaxies & Purple open triangle&\gr\ sources\\
Blue open square&Crates source&Purple filled pentagon&Pulsar\\
Dark green circle&Quasars&Green concave open quadrilateral& GRB \\
\hline\hline
\end{tabular}
\caption{Symbol meaning of the candidate map and Radio-X-ray map}
\label{symbol}
\end{center}
\end{table*}

\begin{table*} [h!]
\begin{center}
\begin{tabular}{lc}
\hline\hline
Color&Waveband\\
\hline
Red&Radio\\
Orange&Infrared\\
Gold&Optical\\
Green&Ultraviolet\\
Blue&X-ray\\
Purple&\gr\,\\
\hline\hline
\end{tabular}
\caption{Color used in the error circle maps}
\label{errormap}
\end{center}
\end{table*}

%% file: results.tex
\section{Examples of the use of VOU-Blazars and some results}
\label{resultvo}
In this section, we illustrate two examples of the use of VOU-Blazars and describe the output.
\begin{figure}[h!]
\begin{center}
\includegraphics[width=1.\linewidth]{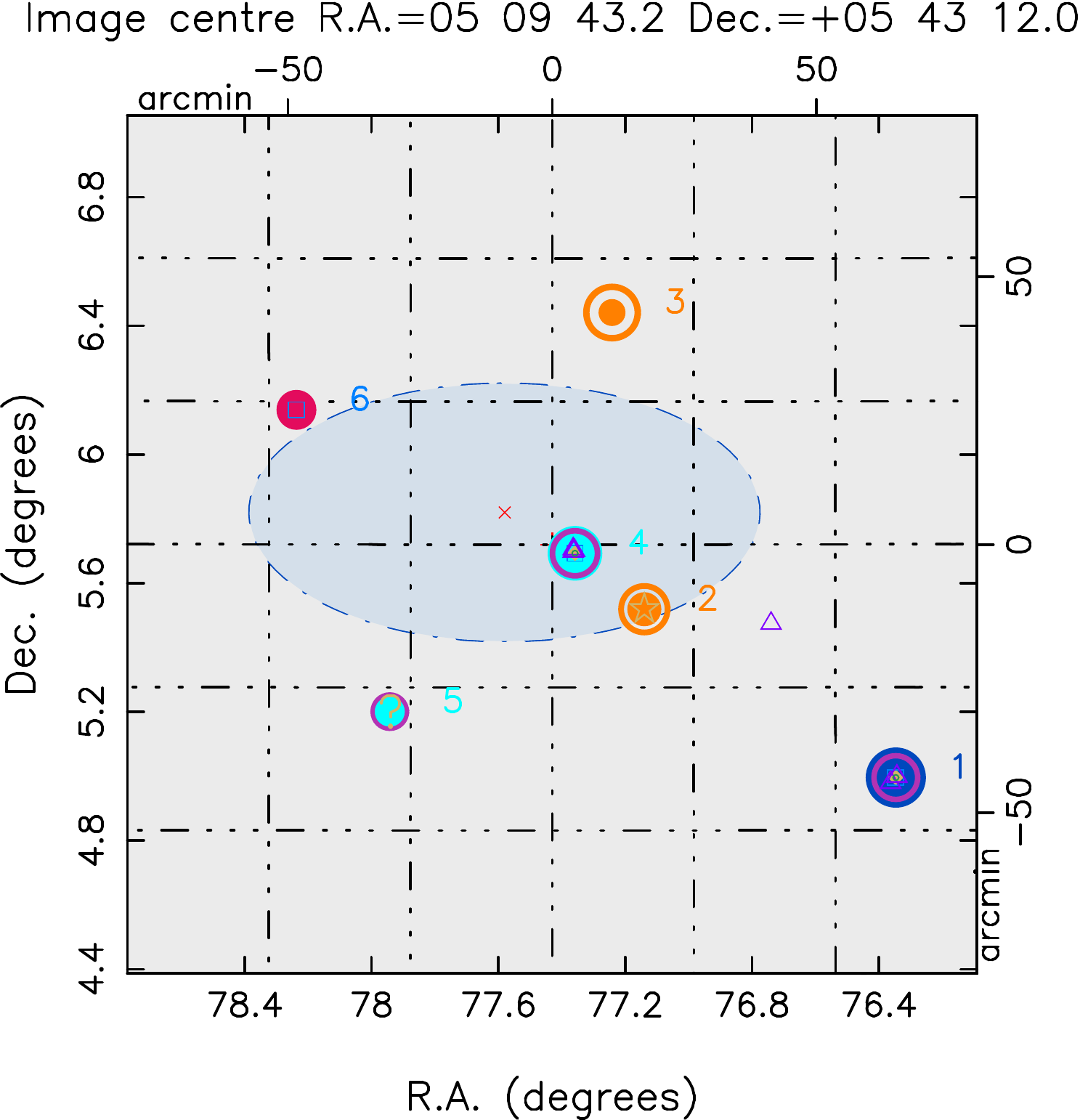}\\
\includegraphics[width=1.\linewidth]{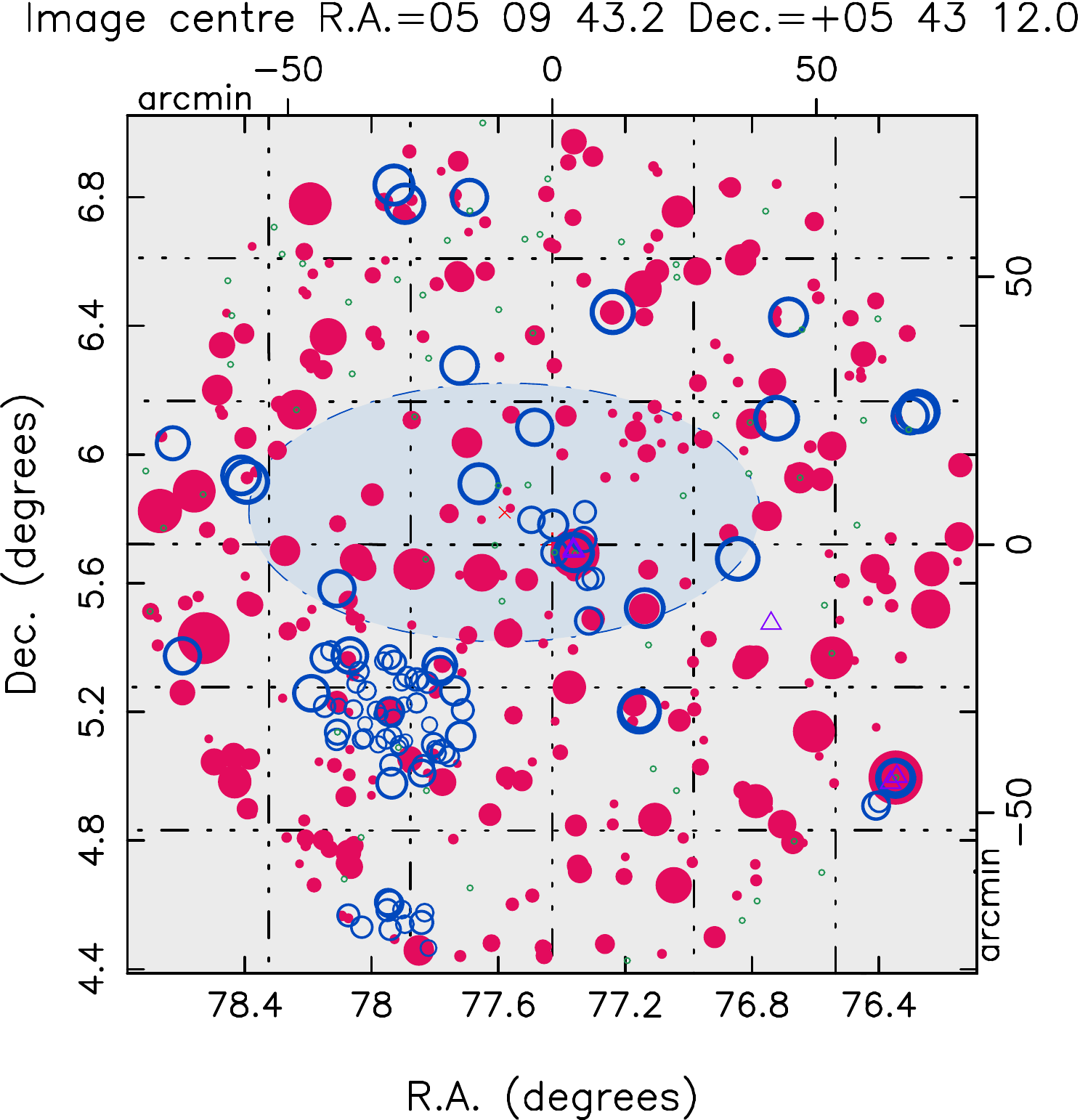}
\end{center}
\caption{%Note that the X-ray data is not uniformly distributed, reflecting the fact that some parts of the map have been observed with sensitive telescopes like XMM or Swift, but for the most part only data from the relatively low-sensitivity RASS survey are available. 
Top: Map showing all blazars candidates found in the area of interest. Orange symbols: HBL/HSP blazars, Cyan: IBL/ISP, Blue: LBL/LSP, the red circle with open square represent a flat radio source (from the CRATES catalog) with no associated X-ray detection. 
Bottom: Map showing all radio (red filled circles) and X-ray (blue open circles) sources in the field of interest. 
Both maps refer to case 1. See text for details.}
\label{outputcand1}
\end{figure}

The simplest way to use the VOU-Blazars tool is to input three parameters: R.A., Dec., and radius, to define the position in the sky and the radius of a circular area where to search for blazars.
The RA and Dec are in the unit of degrees (epoch J2000.0), and the searching radius is in the unit of arc-minutes, as in the following example.

\begin{verbatim}
$ VOU-Blazars 77.43 5.72 80 
\end{verbatim}

Additional parameters, such as the value of NH (Hydrogen Column density in the Galaxy at the requested position, in units of ${\rm cm}^{-2}$), and the parameters to define up to two circular (or elliptical) search areas can also be specified: 
%For example, the tool could be run with the following commands
 
\begin{verbatim}
$ VOU_Blazars 77.43 5.72 80 3.e21 50 60 25 45
\end{verbatim}
The value of NH is used to de-absorb the (near-IR, optical or X-ray) flux values retrieved from the catalogs to take into account of absorption in our galaxy. 
If no NH value is provided, VOU-BLazars uses the NH value obtained from the NH tool within HEASOFT\footnote{https://heasarc.gsfc.nasa.gov/docs/software/heasoft/}. 

The first case considered above refers to a search in a circular region centered at the position R.A. = 77.43 and Dec. = 5.72 degrees
%of the astrophysical neutrino IceCube-170922A 
%\citep{Kopper2017}, 
with a radius of 80 arc-minutes. 
In the second example the input further specifies the nH value (3.e21 cm-2), one circular region (with radius 50 arcmin), and one elliptical area (major and minor axis of 60 and 25 arcmins and position angle (45 degrees). 
See the GitHub page for more information on input parameters (see Sect.\ref{availability}).

At the end of the first phase, the tool produces two figures, one candidate map and one Radio-X-ray source map, that are shown in the top and bottom parts of Fig. \ref{outputcand1}. The meaning of the symbols in the source map is explained in Table~\ref{symbol}. 
The circular area of radius 50 arcmin is where sources are searched, while the elliptical area is plotted in light blue color, which in this example approximates the uncertainty in the arrival direction of the neutrino IC170922.
%and the smaller area is set to 15 arcmins.
%{\color{red} do we really need the 15 arcmin circle?)}
%for testing. 

Six known or candidate blazars are found within the searching region. 
Source number 4, the closest to the center, is the bright blazars TXS 0506+056 that has been associated to the neutrino \citep{Icecube2018a,Padovani2018}. It is an ISP blazar and therefore it is represented with a light blue circle.
%also cataloged in 5BZCat (5BZBJ0509+0541) and CRATES (J050926+054143) with golden diamond and blue square on the figure.
Two HBL candidate blazars (orange circles) are also found around the neutrino detection, source number 2, inside the 90\% error ellipse, is a  3HSP source (3HSP J050833.3+053109) and it is also marked with golden star); source number 3, outside the error ellipse, is an uncatalogued HSP candidate. 
Source number 1, an LSP object (dark blue circles) is a known source listed in 5BZCat (5BZQ J0505+0459), and source number 5 is an ISP candidate close to cluster of galaxies (ZW 4472), and for this reason it appears as 
a light blue symbol with a question mark. 
In addition, there are two \gr\ detections (3FHL J0505.4+0458 and 4FGL J0505.3+0459), represented by purple triangles.
Source number 6, the flat-spectrum radio source CRATESJ051256+060835, is without an X-ray counterpart and is shown as the red filled circle (radio source) with a blue open square (CRATES sources) on it. 
Source numbers 1 and 4 are also in million quasar catalog with small dark green circle on it.
Note that for candidate numbers 1, 4, and 5, the size of the X-ray counterpart on the candidate map is smaller than that of the radio counterpart, causing the open circle to be covered by the filled circle.
In these situations, the X-ray counterparts are shown with a magenta color. 

\begin{figure} [h!]
\begin{center}
\includegraphics[width=0.95\linewidth]{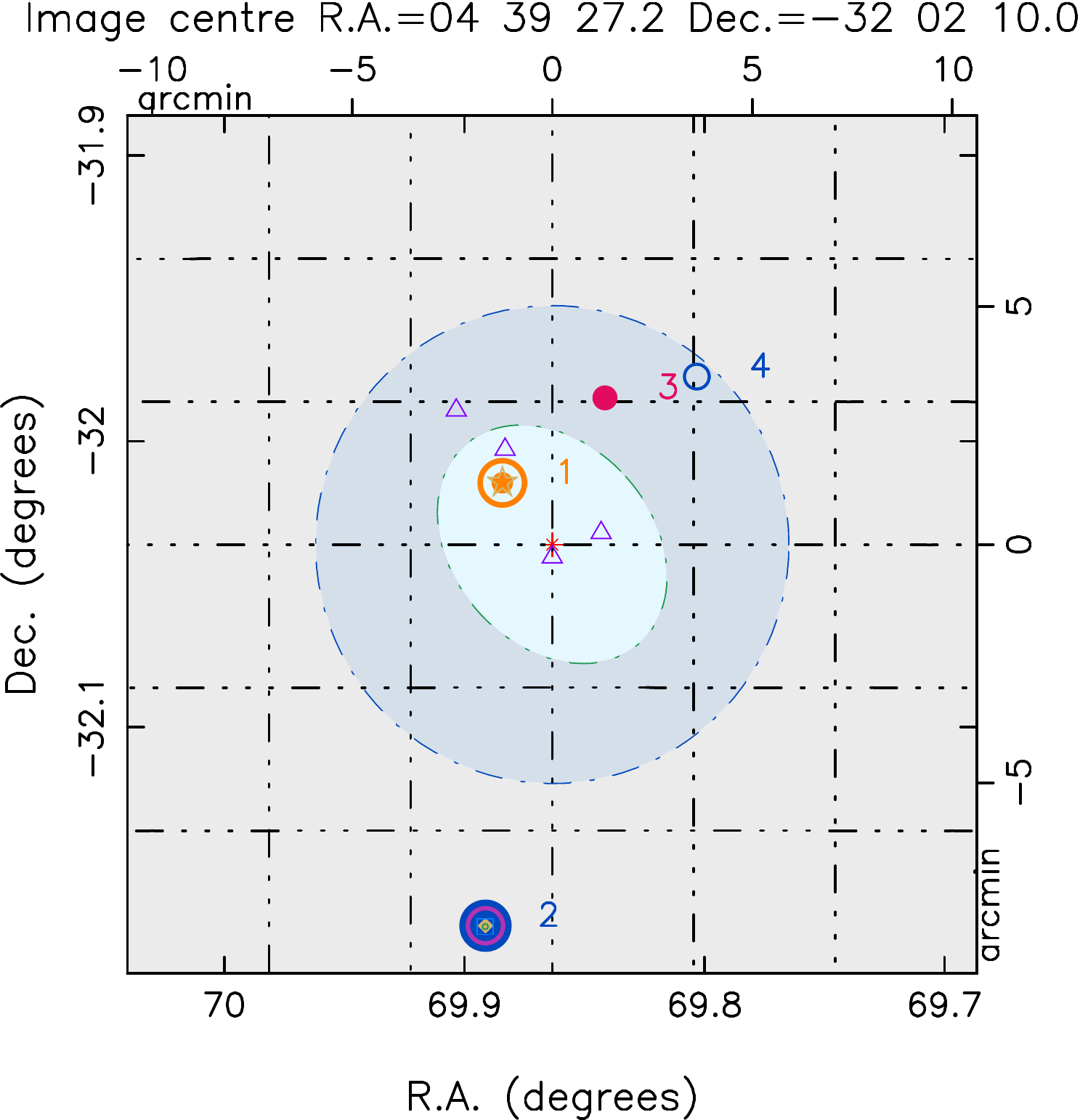}
\includegraphics[width=0.95\linewidth]{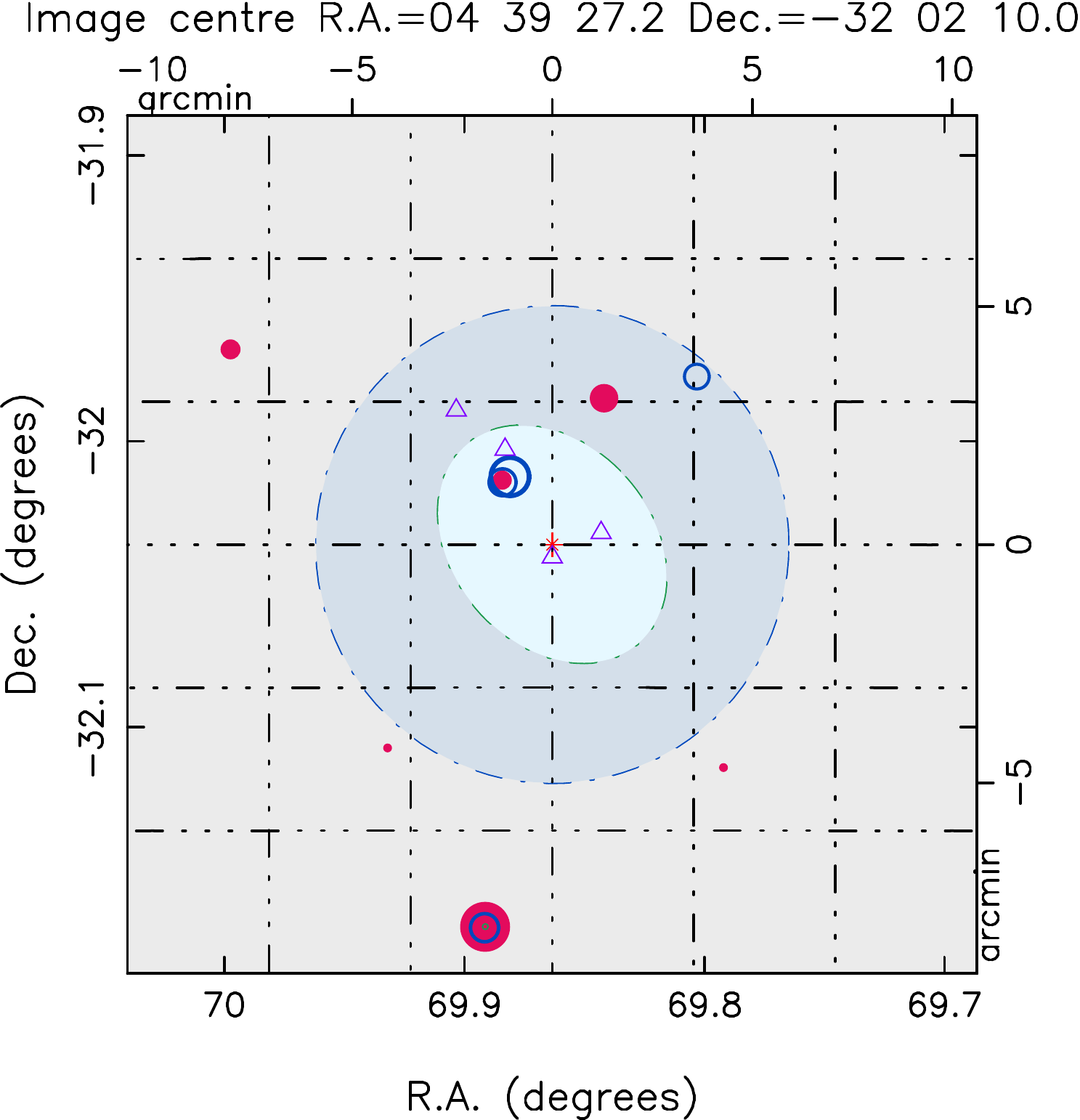}
\end{center}
\caption{Same as fig.\ref{outputcand1} for case 2. See text for details.}
\label{outputcand3}
\end{figure}

In the example illustrated by Fig. \ref{outputcand3} the VOU-Blazars tool was run on the position of the {\it Fermi} \gr\ source 4FGL J0439.4-3202.
%within 40 arcmins. 
The elliptical shape represents the uncertainty on the arrival direction of the \gr\ source, while the larger area is a circle with radius of 5 arcminutes, approximately half the size of the total searching radius.
%and double of the \gr\ position uncertainty. 
%We added this larger error circle to ensure not to lose possible blazar candidates just outside the \gr\ detection. 
The candidate map shows two known blazars: the HSP 3HSP J043932.2-320052 and the LSP 5BZU J0439-3210, which is also in CRATES (J043929-320956).
%within the specified radius. 
The 3HSP source, candidate number 1, is inside the error region of the {\it Fermi source} and is very likely the counterpart of the \gr\ source. 
Sources numbered 3 and 4 are also plotted in the candidate map and listed as possible blazars, following the comparison of their positions and fluxes with 4.8GHz and UV catalogues data in the intermediate phase.
%found and may be worth checking as well. 
%These two sources without radio-X-ray matched are found during the . 
%Source number 3, within larger error circle 

\begin{figure} [h!]
\begin{center}
\includegraphics[width=0.9\linewidth]{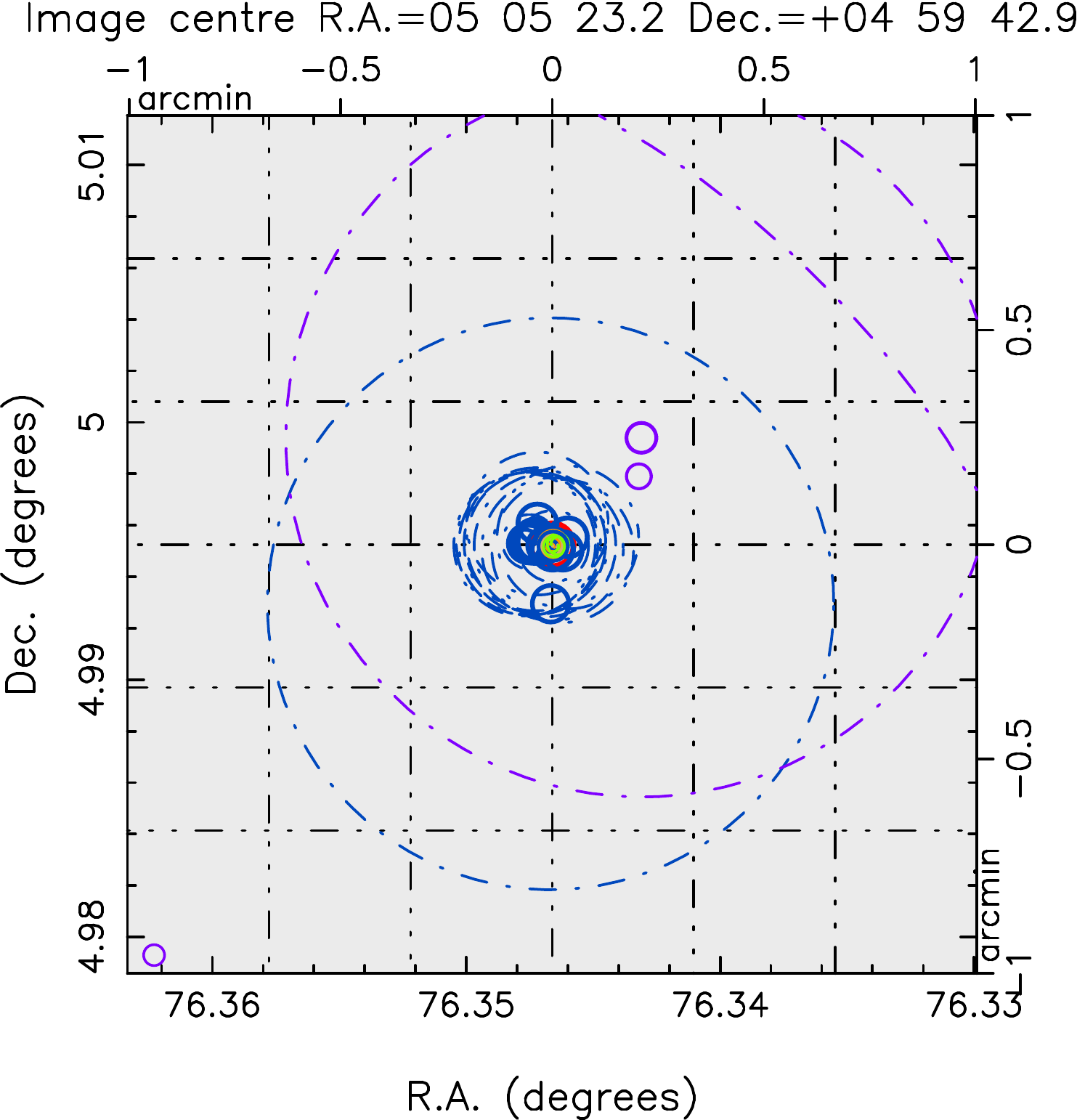}\\
\includegraphics[width=1.0\linewidth]{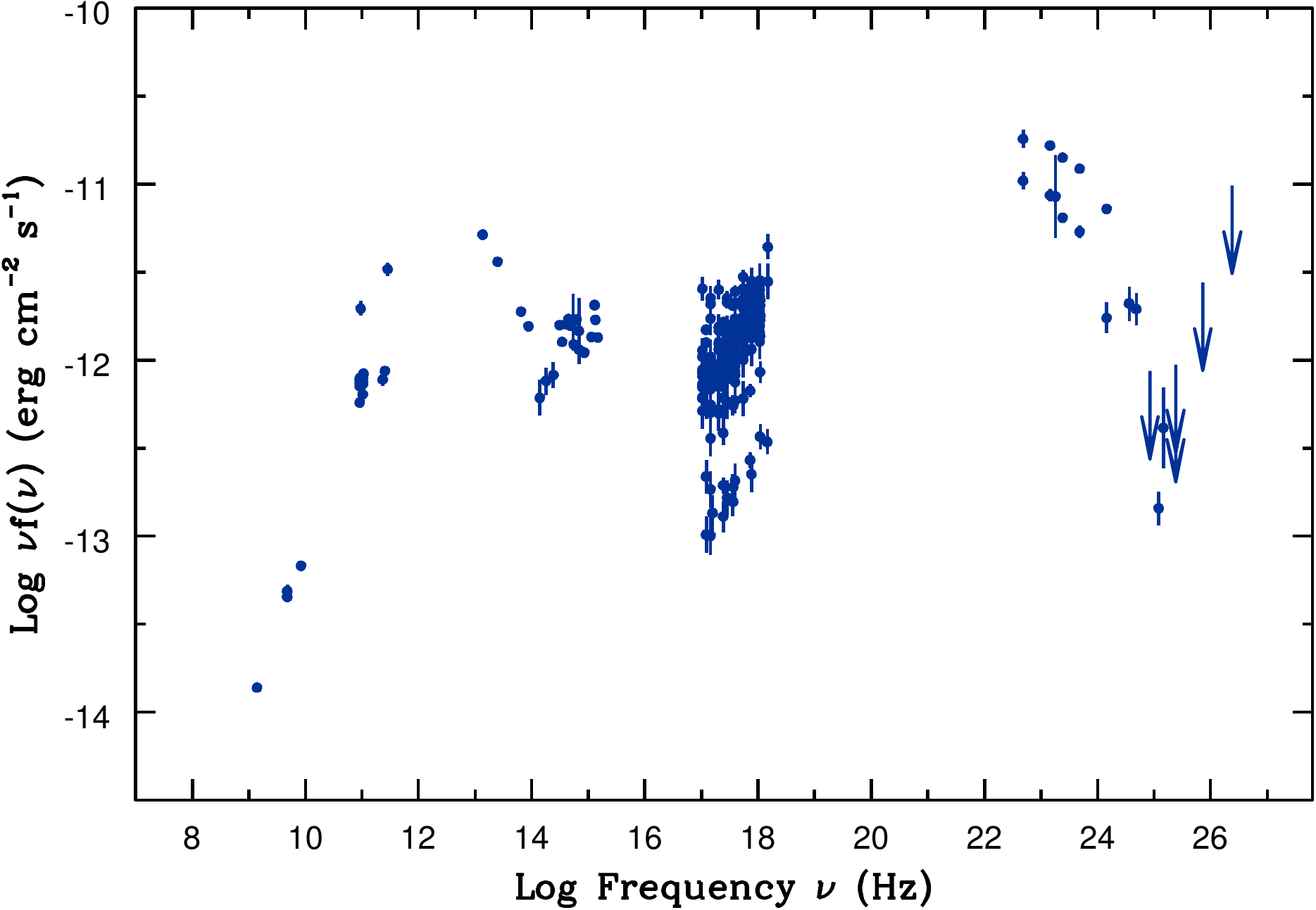}
\end{center}
\caption{Top: Error circle map showing the error region of X-ray (blue circles), \gr\ (purple circles) and IR/optical/UV (orange,green circles) sources in the area where the multi-frequency data are collected. Bottom: The SED of candidate number 1 of case 1 (Fig.\ref{outputcand1}, an LSP blazar). The color blue corresponds to the color coding for LBL/LSP blazar candidate shown in Fig.\ref{outputcand1}}
\label{sederror}
\end{figure}

\begin{figure} [h!]
\begin{center}
\includegraphics[width=0.9\linewidth]{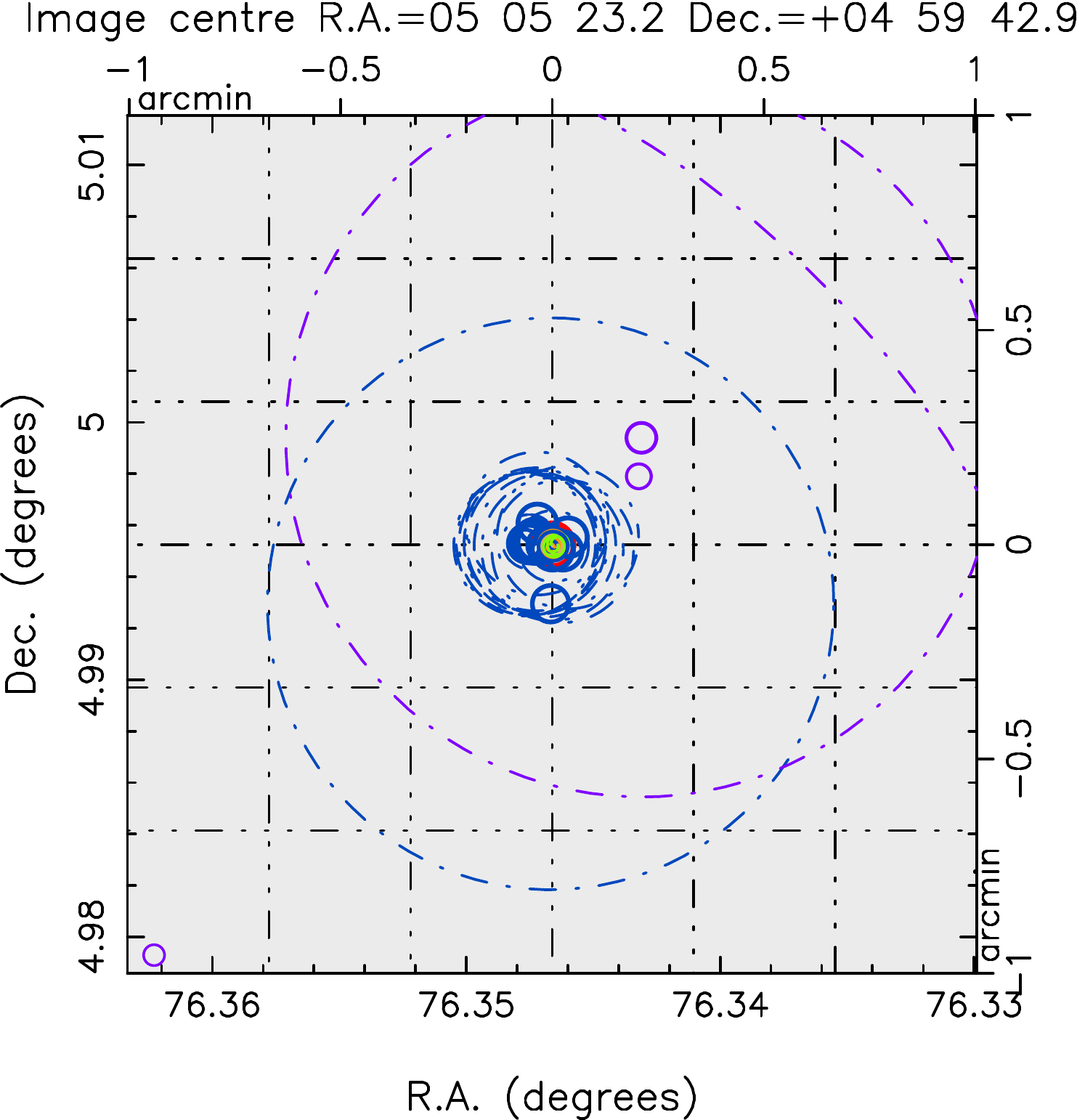}\\
\includegraphics[width=1.0\linewidth]{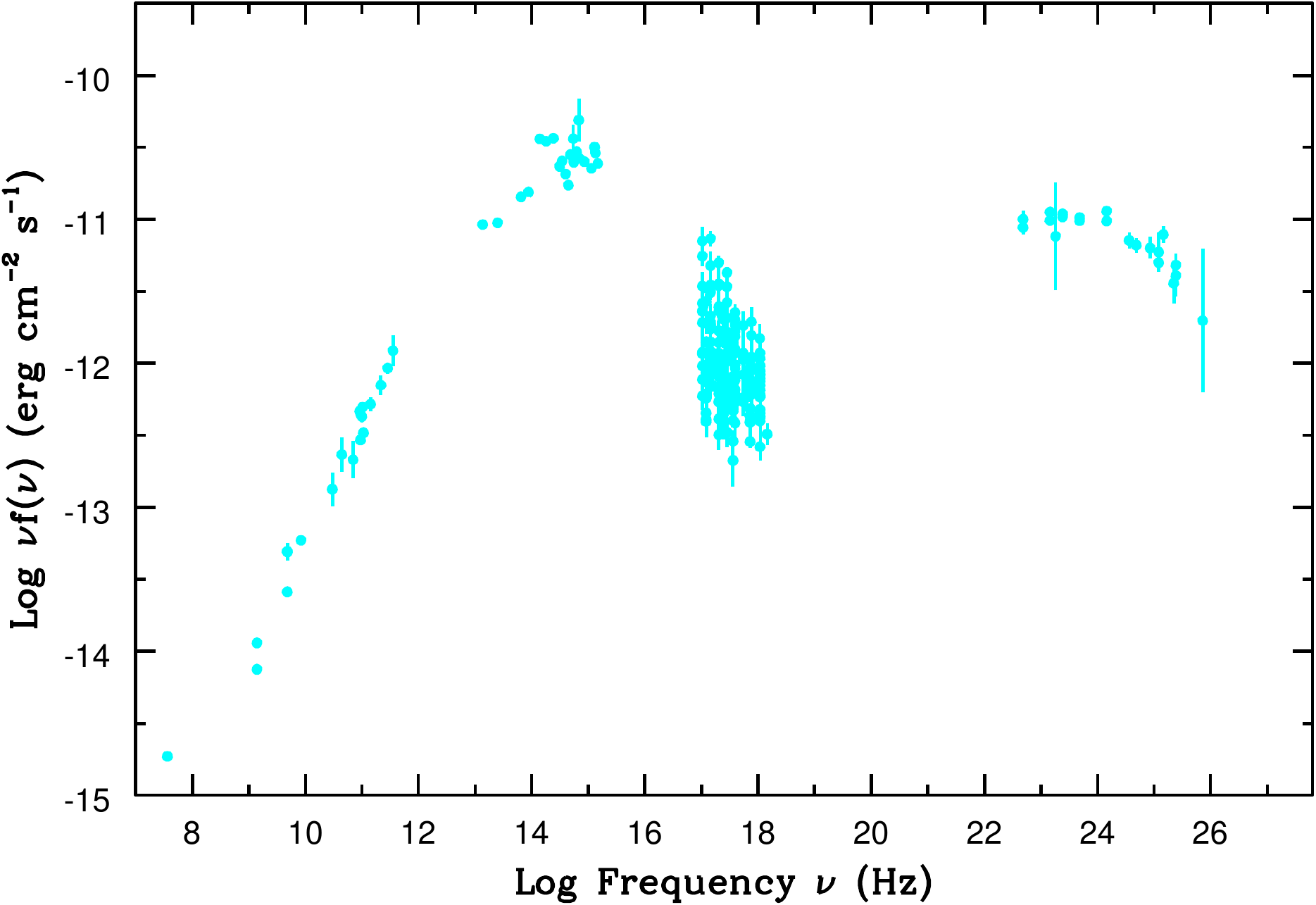}
\end{center}
\caption{Same as Fig.\ref{sederror} for candidate number 4, an ISP, in case 1}
\label{sederror2}
\end{figure}

\begin{figure} [h!]
\begin{center}
\includegraphics[width=0.9\linewidth]{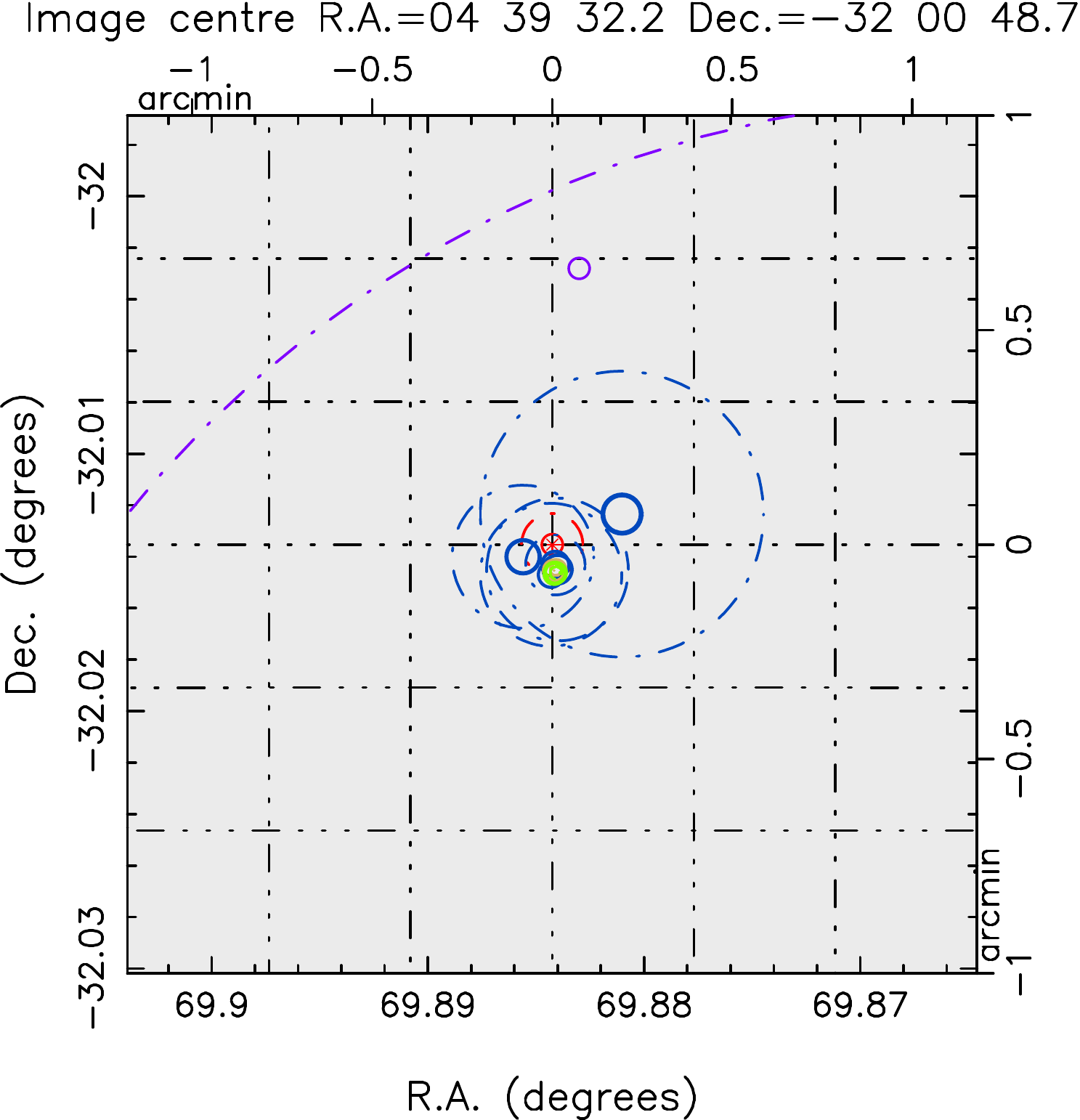}\\
\includegraphics[width=1.0\linewidth]{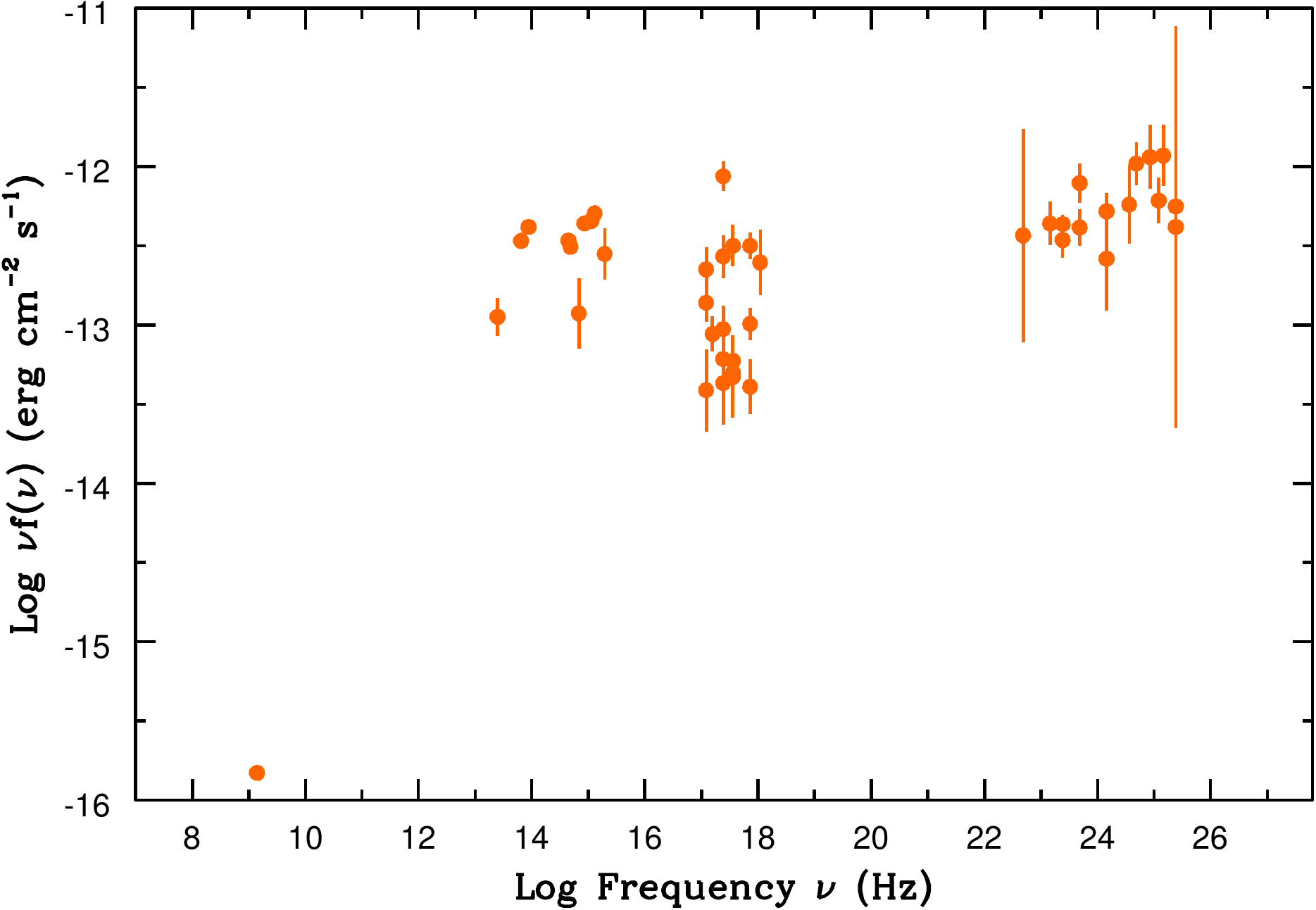}
\end{center}
\caption{Same as Fig.\ref{sederror} for candidate number 1, an HSP, in case 2}
\label{sederror3}
\end{figure}

After the second phase, the VOU-Blazars plots the SED of the requested candidate(s) and the associated error circle map(s).  Fig.~\ref{sederror}, \ref{sederror2}, and \ref{sederror3} are examples of these plots. %referring to candidate number 1 (LSP) and candidate number 4 (ISP and possible neutrino counterpart, TXS 0506+056) and candidate number 1 in case 2 (3HSP).
The meaning of the color of each error circle is given in Table~\ref{errormap}.

\subsection{The VOU-Blazars SED mode}
\label{sedmode}
The VOU-Blazar tool can also be used to directly produce SEDs of objects with known position. 
This is called the "SED mode" and can be activated from the command line, as in the following example 
\begin{verbatim}
$ VOU-Blazars RA DEC r s 
\end{verbatim}
where "r" is the radius of a circular area, typically 1 arcminute, where to plot the error uncertainties of the data retrieved, and "s" is a flag specifying that the tool should run in SED mode.
The same mode can also be activated from the Open Universe portal by clicking on the icon indicated by the blue dashed arrow in Fig. \ref{ou_portal}.

\begin{figure*} [h!]
%\hspace{-0.8truecm} 
\includegraphics[angle=0,width=1.0\linewidth]{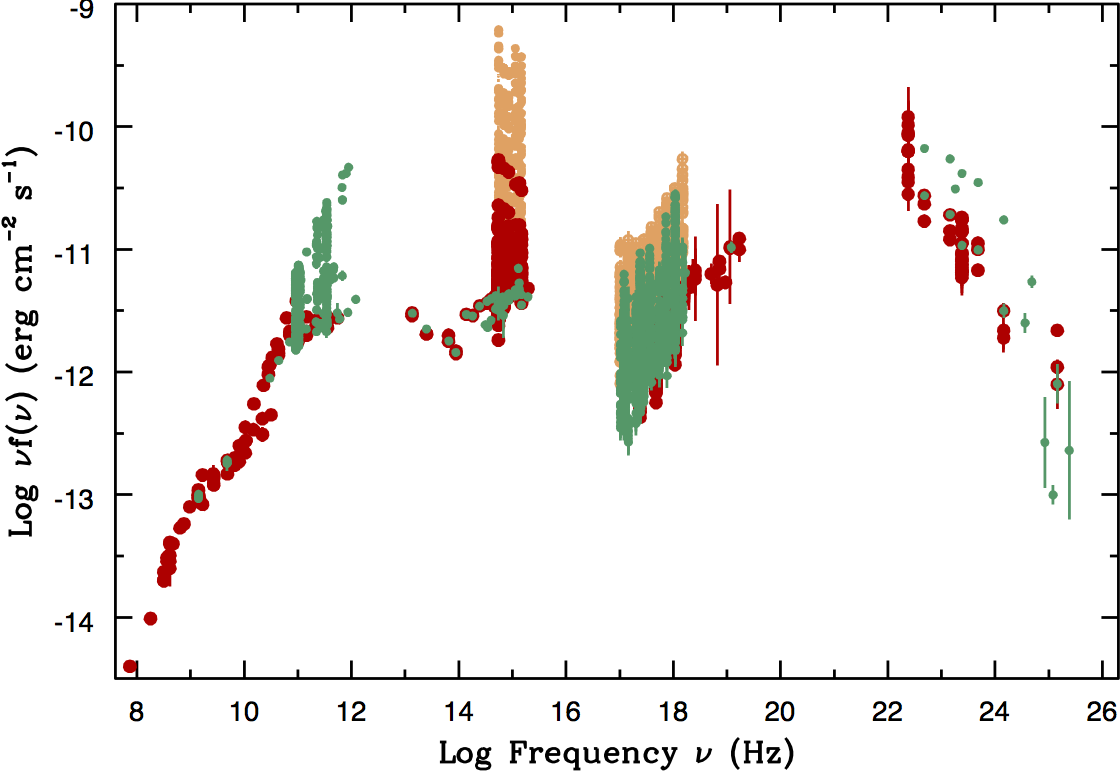}
\caption{The SED of the blazar CTA 102 obtained combining the data from VOU-Blazars (green color) and those directly retrievable from the SSDC SED tool (dark red) or open data not available through the VO and manually loaded on the SSDC tool (light brown).}
\label{cta102}
\end{figure*}

\begin{figure*} [h!]
\includegraphics[angle=0,width=1.0\linewidth]{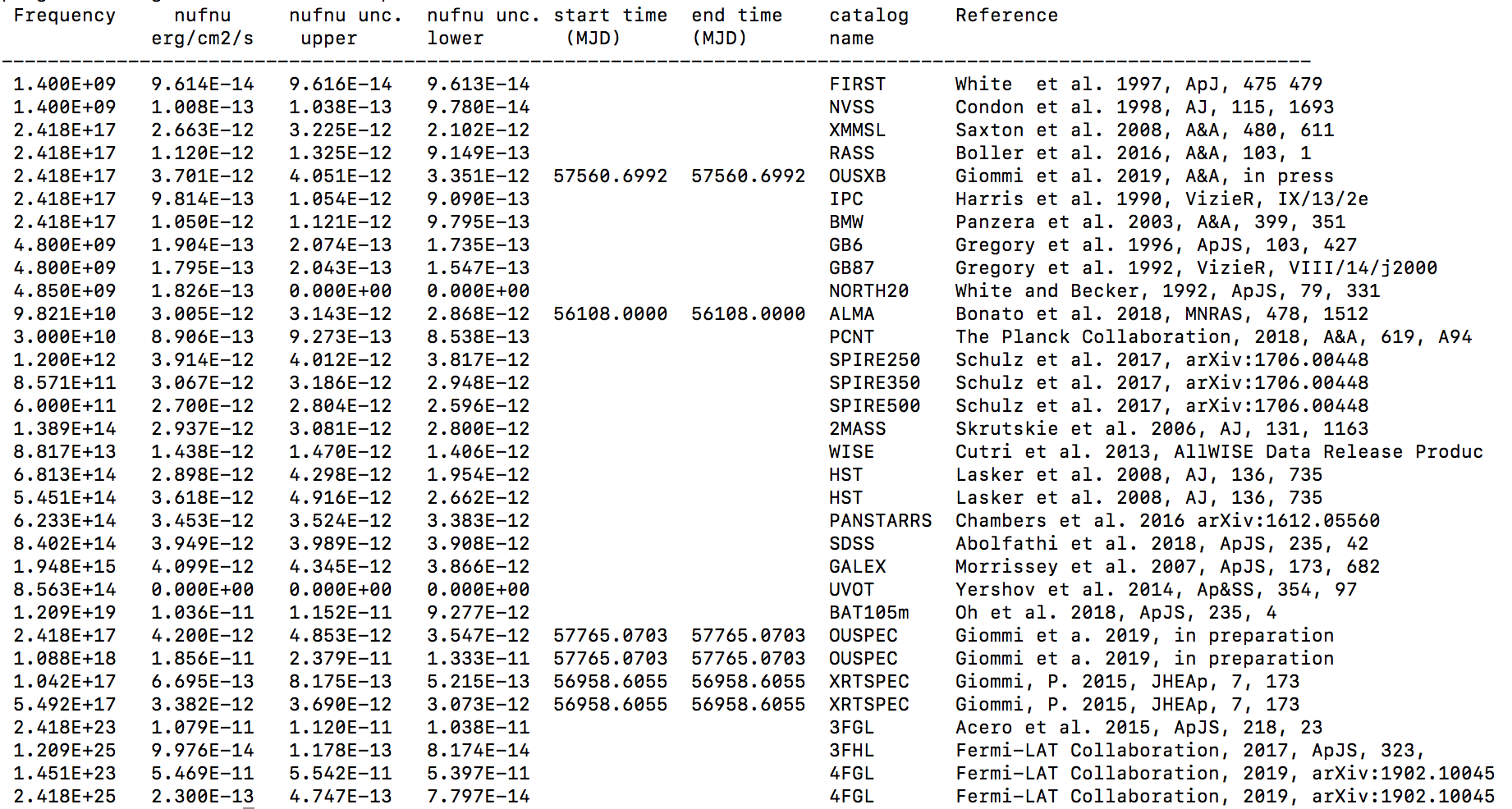}
\caption{The file containing the $\nu$f($\nu$) flux, the corresponding uncertainties, observation date (when available) and the bibliographic references used by VOU-Blazars to build the SED of CTA102 shown in Fig.\ref{cta102}. For reasons of brevity only one entry per catalog or database and per representative frequency are reported. The real file includes over 1600 lines.}
\label{cta102SedTable}
\end{figure*}

The SED mode of VOU-Blazars is designed to build the SED of a source with known precise coordinates.
%without going through the processes of finding blazar candidates. 
The tool finds all the counterparts from every catalogs of the first and second phase without interruption.  
Comparing with the usual Find Candidate mode, the SED mode only returns flux and error circle along with the SED and error circle map for all the matched counterparts rather than returning a list and a map of potential candidates. 
%Thus, the user does not need to further specified an interested candidate during the working process. 
Cross-matched radius in SED mode is the position error of each searching catalogs. 
Fig.~\ref{cta102} shows an example of the SED generated by VOU-blazars compared to the one produced by Italian Space Agency (ASI) Space Science Data Center (SSDC) SED Builder tool\footnote{https://tools.ssdc.asi.it/SED/}. 
The corresponding file containing flux values and references is illustrated in Fig.~\ref{cta102SedTable}.
The data on SED obtained from the VOU-Blazars are very similar to that of the SSDC tool. 
Moreover, the VOU-Blazars SED contains a number of measurements that the SSDC tool did not retrieve. 
On the other hand the SSDC SED tool shows many other points in different colors, some of which were retrieved from sites not reachable via VO protocols and uploaded manually. 
The two tools are therefore complementary.
%{\bf We note that the SED data file is output in various formats for different purpose. A CSV type file is used for TOPCAT }

\newpage

%% file: application.tex
\section{Previous applications of the tool}

A preliminary version of VOU-Blazars has already been used in some scientific papers. 
In particular, the tool was extensively used by some of us to find new HSP blazars to be included in the 3HSP catalog\citep{Chang2019}. 
\citet{Padovani2018} used the tool to study in detail all the possible counterparts of the neutrino event IceCube-170922A \citep{Icecube2018b}.
The tool has been implemented as one of the options available within the Open Universe portal, where it can be used also to identify sources with large positional uncertainty. Figure \ref{ou_portal} shows the case of the source 4FGL J1211.6+3901. %which was obtained by inserting the name of the Fermi-LAT source in the input area in the red box and by clicking on the VOU-Blazars icon highlighted by the red arrow.
%in could successfully recognize all the possible counterparts of a VHE source within a relatively large area (REFERENCE). 

Figure~\ref{outputcand1} is the candidate map which lists all the possible counterparts for IceCube-170922A. 
The larger error elliptical is 90 \% of the neutrino position, and the smaller error circle is set to 15 arc-minutes for testing. 
Source number 4, and ISP, TXS 0506+056, also cataloged in 5BZCat (5BZBJ0509+0541) and CRATES (J050926+054143) with the blue square and golden diamond on the figure.
This source is the closet and most possible counterpart for the neutrino event.\citep{Icecube2018a,Padovani2018} 
The ISP is also detected by {\it Fermi} and in 3FHL catalog (3FHL J0509.4+0542), with purple triangles nearby. 

Apart from identifying possible counterpart for VHE/neutrino detections, there were many good HSPs found by running the tool.
Specifically, around 30 new sources added to the 3HSP catalog are selected with the VOU-Blazars. 
The tool retrieved data from more catalogs (GAIA, PanSTARRS, XMMOM) or the latest version of some catalogs (GALEX, XMM, Swift XRT) to plot the SED than the other SED tool. 
Thus, the SED built by VOU-Blazars may contain more data and could identify more HSP candidates. 
With more data, the synchrotron peak value could be refined as well.

\section{Availability of the tool}
\label{availability}

In compliance with the \OU\ principles of transparency and easiness of use, the VOU-Blazars tool is openly available in various forms, as described below.

\subsection{Source code on GitHub}

The VOU-Blazars source code can be downloaded from GitHub at \url{https://github.com/ecylchang/VOU\_Blazars}.
To install the VOU-Blazars code and run the tool on your own computer, please follow the instructions available at the the GitHub page. 

\subsection{Implementation as a Docker container}

A version of the VOU-Blazars tool encapsulated in a Docker 
container is available at the following address \url{https://hub.docker.com/r/chbrandt/voublazars}, which allows the user to run VOU-Blazars in a simple way.

Because Docker command-lines may be complex depending on the user's familiarity to such environment we developed a small Python tool that simplifies the use of Docker for VOU-Blazars.
The tool is called \texttt{dockeri} and is available at \url{https://github.com/chbrandt/dockeri}. \texttt{Dockeri} works with \textit{any} version of Python. Once Docker\footnote{\url{https://docs.docker.com/install/}} and \texttt{dockeri}\footnote{\url{https://github.com/chbrandt/dockeri}} are installed,  VOU-Blazars is run with a command-line like:
\begin{verbatim}
$ dockeri -w output_dir \
         chbrandt/voublazars
\end{verbatim}
, which will run the latest version of VOU-Blazars.

If no positional arguments are given -- like the above example --, VOU-Blazars will ask for them interactively.
The user can also directly give the input parameters:
\begin{verbatim}
$ dockeri -w output_dir \
         chbrandt/voublazars \
         77.43 5.72 1 s
\end{verbatim}
, which runs VOU-Blazars in \textit{SED mode} (see section~\ref{sedmode}).
After the processing is complete, the results will be found in directory \texttt{output\_dir}.

\subsection{Implementation within the Open Universe portal}
\begin{figure*} [h!]
%\hspace{-0.8truecm} 
\includegraphics[angle=0,width=1.0\linewidth]{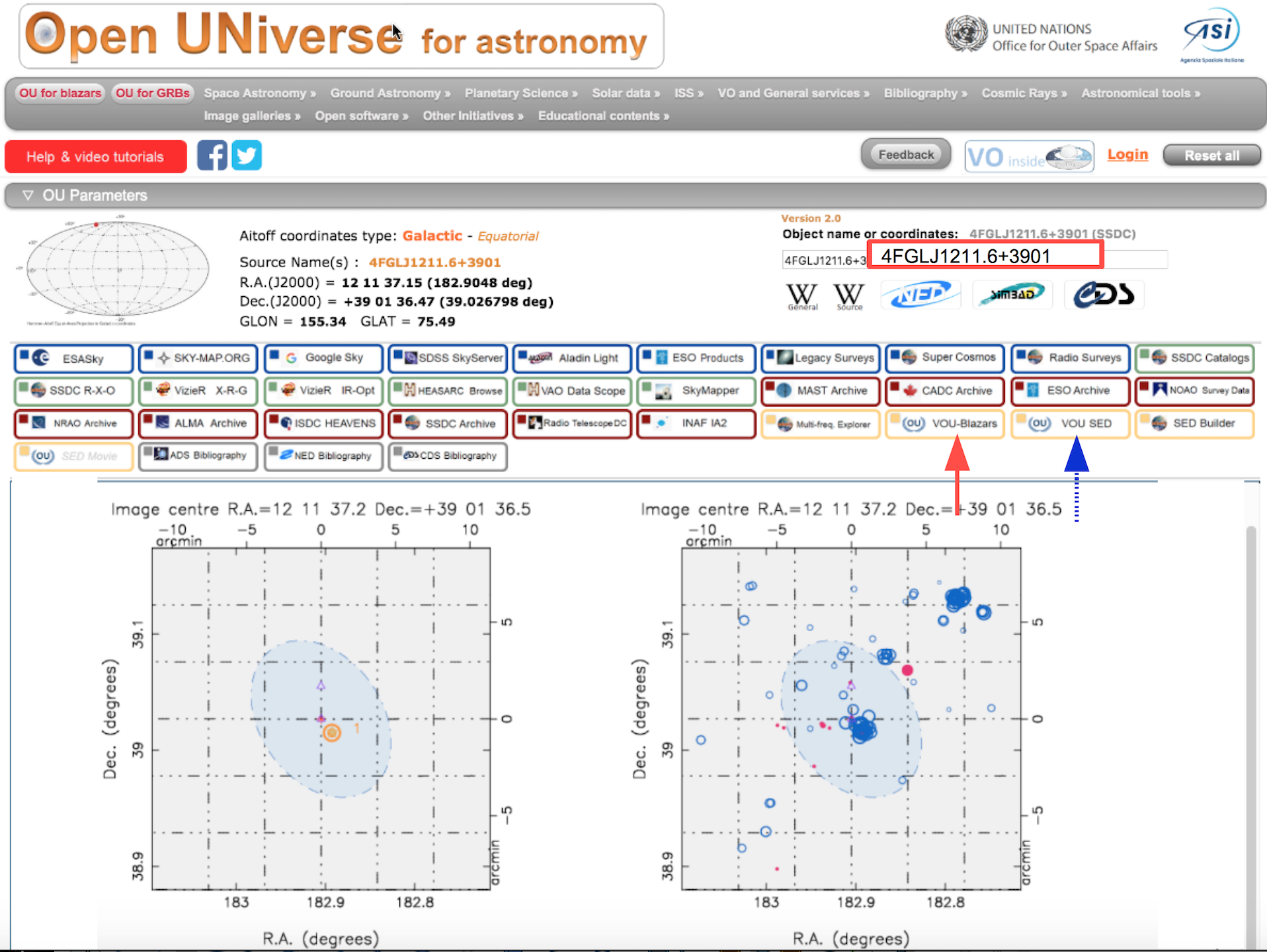}
\caption{The implementation of VOU-Blazars within the Open Universe portal. This example shows the case of the Fermi $\gamma$-ray source 4FGLJ1211.6+3901, whose position and elliptical uncertainty region parameters have been retrieved by the system. The candidate blazars map (bottom left) showing the blazar 3HSP J121134.2+390053 in orange color to be the only likely counterpart in the 95\% error ellipse, and the radio/X-ray sources map (bottom right) has been produced by VOU-Blazars by clicking on the icon pointed by the red arrow.}
\label{ou_portal}
\end{figure*}

VOU-Blazars has been integrated with the \OU\ web portal. 
%as of the options othat are activated when the name of an astronomical object, or its sky coordinates, is entered in the input area (see Fig. \ref{ou_portal}). 
To use this on-line version of the tool simply enter the desired sky coordinates (or the name of an astronomical object) in the input area located on the top right side of the portal, highlighted by a red box in Fig. \ref{ou_portal}, then click on the VOU-Blazar button indicated by the red arrow. The figure gives the example of the region around a Fermi 4FGL $\gamma$-ray source. For sources from catalogues of objects that have non negligible positional uncertainty the Open Universe portal retrieves both the best coordinates of the source and the parameters of the uncertainty region.   

%Fig. \ref{ou_portal} gives an example of usage of the 
%tool within the \OU\ portal, for the case of the IceCube
%high energy astrophysical neutrino that has been associated %to the blazar TXS0506+056 \citep[Add Reference]{IceNeutrino}.

%\begin{figure*} [h!]
%\hspace{-0.8truecm} 
%\includegraphics[angle=-90,width=1.2\linewidth]{figures/VOU-portal.pdf}
%\caption{The implementation of VOU-Blazars inside the Open Universe portal at openuniverse.asi.it. The example shows the case of the IceCube neutrino detected on 22 September 2017. The candidate blazars map and the radio/X-ray sources map have been produced by VOU-Blazars by clicking on the icon pointed by the red arrow.}
%\label{ou_portal-}
%\end{figure*}

\newpage

%% file: conclusion.tex
\section{Conclusion and future developments}
%Some VHE/neutrino detections have not been associated with any lower-frequency sources yet. 
We have developed VOU-Blazars, a new software tool that makes use of multi-frequency data-sets retrieved using VO protocols, to find blazars anywhere in the sky, and particularly in moderately large areas (up to a few degrees), that are typical of the positional uncertainty of $\gamma$-ray sources, high-energy astrophysical neutrinos, UHECRs,  etc.

%The first version of the VOU-Blazars can be downloaded from the link \url{https://github.com/ecylchang/VOU_Blazars}. 
%With the tool, all the interesting radio or X-ray sources with UV counterparts or Gamma-ray counterparts may not be missed anymore as it could return good blazars candidates from all available lists of radio and X-ray emitters after dedicated examination. 

During the development and testing phase the VOU-Blazars tool has been used to find potential IceCube neutrinos counterparts \citep{Padovani2018}, and to discover several new HSP blazars during the preparation of the 3HSP catalog\citep{Chang2019}. 
The consolidated version of VOU-Blazars presented in this paper will be useful to discover new blazars and for the identification of the counterparts of the many high-energy extragalactic sources discovered by the current and by the next generations of \gr\, and neutrino observatories.

%We plan to further develop the tool and provide the updated versions on-line.
%Currently, the VOU-Blazars initially classifies the sources based on their radio-X-ray slope flux ratio.
In the near future, we plan to make a number of improvements to the tool, including
\begin{itemize}
    \item enhanced identification methods based on more multi-frequency and multi-temporal information; 
    \item the use of machine learning techniques;
    \item increase the number of catalogs providing more spectral and timing information
\end{itemize}

All of the above will keep the tool up to date with new catalogs as they come on-line, will make it more precise in the identification process
(reducing the number of uncertain candidates that may generate confusion in large searching area), and will enable spectral and timing analysis. 

%Some catalogs, like OUSXB, ALMA, and VERITAS provide timing information with VOU-Blazars, and we will make the light curve display more sophisticated. 
%There were more catalogs such as and VLA, VLASS...etc, could be applied to the tool to make the searching more complete and have more data on SED.
%With all the above modifications, there will be an updated and better version of VOU-Blazars in near future. 

\newpage

%% file: acknowledgment.tex
\section*{Acknowledgment}
\textbf{YLC} is supported by the Government of the Republic of China (Taiwan). Most of this work was carried out at Agenzia Spaziale Italiana, as part of the Open Universe initiative, and at the University La Sapienza of Rome, Department of Physics. We make use of archival data and bibliographic information obtained from the NASA/IPAC Extragalactic Database (NED), data and software facilities from the ASDC managed by the Italian Space Agency (ASI).   

\textbf{CHB} acknowledges the support of ICRANet and the Brazilian government, funded by the CAPES Foundation, Ministry of Education of Brazil under the project BEX 15113-13-2.

\textbf{PG} acknowledges the support of the Technische Universit\"at M\"unchen - Institute for Advanced Study, funded by the German Excellence Initiative (and the European Union Seventh Framework Programme under grant agreement no. 291763)

We thank Sara Turriziani and Theo Glauch for helpful suggestions.

%% file: reference.tex
\section*{References}
\bibliographystyle{elsarticle-harv} 
\bibliography{voblazar.bib}